# Experimental and Theoretical Investigation of the Barrier Pyroelectric Effect in a Quantum Paraelectric Semiconductor

A. V. Butenko[a], V. Sandomirsky[a], R. Kahatabi[a], Z. Dashevsky[b], V. Kasiyan[b], Z. Zalevsky[c] and Y. Schlesinger[a]

[a] Department of Physics, Bar-Ilan University, Ramat-Gan 52900, Israel
[b] Materials Engineering Department, Ben-Gurion University of the Negev, P.O.B. 653, Beer-Sheva 84105, Israel
[c] School of Engineering, Bar-Ilan University, Ramat-Gan 52900, Israel

## ABSTRACT

We describe here the first comprehensive investigation of a pyroelectric response of a *p*-*n* junction in a non-polar paraelectric semiconductor. The pyroelectric effect is generated by the, temperature dependent, built-in electrical dipole moment. High quality PbTe *p*-*n* junctions have been prepared specifically for this experiment. The pyroelectric effect was excited by a continuous $CO_2$ laser beam, modulated by a mechanical chopper. The shape and amplitude of the periodic and single-pulse pyroelectric signals were studied as a function of temperature (10 K – 130 K), reverse bias voltage (up to -500 mV) and chopping frequency (4 Hz – 2000 Hz). The pyroelectric coefficient is $\approx 10^{-3}$ μC/cm$^2$K in the temperature region 40 - 80 K. The developed theoretical model quantitatively describes all the experimental features of the observed pyroelectric effect. The time evolution of the temperature within the *p*-*n* junction was reconstructed.

## I. INTRODUCTION

In a recent Letter [1] we reported the first experimental verification of the existence of a new class of pyroelectric systems, namely, the barrier pyroelectrics, or the barrier pyroelectric effect (BPE). The classical pyroelectrics are crystalline dielectrics of specific symmetry, giving rise to a spontaneous polarization in the absence of an external electric field [2]. One can, however, create a local dipole moment in any solid by a built-in electric field [3]. If this dipole moment depends on temperature, then the corresponding structure will show a pyroelectric response. Such situation arises in semiconductor barrier structures: *p-n* junctions (PNJ), Schottky contacts, heterojunctions etc. The contact potential difference across the barrier, gives rise to a space charge region at the barrier interface [4]. As a whole, this region is neutral, i.e. one has a local dipole moment oriented normally to the interface. If this dipole moment depends on temperature, then a change of temperature brings about a change of the dipole moment, this giving rise to the appearance of a displacement current. This is the barrier pyroelectric effect BPE. In equilibrium, the depolarizing electrical field of the dipole moment is screened by free carriers of the quasi-neutral, outer boundaries surrounding the junction region, similar to a standard pyroelectric.

While a uniform volume density of electric dipole characterizes the classical PE, the BPE is characterized by a dipole moment of the barrier as a whole.

The temperature dependence of the dipole moment can be, in general, attributed to a number of sources: the temperature dependence of the contact potential difference, thermal expansion, temperature dependence of the semiconductor gap, temperature dependence of the lattice dielectric constant ($\varepsilon$), the thermal excitation of carriers etc. Presumably, the strongest temperature dependence of the built-in dipole moment can be reached in substances having a strongly temperature dependent dielectric permittivity, such as the quantum paraelectrics, e.g. $SrTiO_3$ [5] or the narrow-gap semiconductor PbTe [6, 7] – the system investigated in this work.

PbTe is a prospective material for photovoltaic IR sensors, and preparation technology of PbTe PNJs has been previously reported [8]. For the present work we prepared high-quality PbTe *p*-*n* junctions [9, 10], specifically for the investigation of BPE.

The pyroelectric response is observed also in polymer electrets [11, 12]. The BPE is closest to the case “C” in polymers ("Piezo- and pyroelectricity due to heterogeneity of the film and embedded charges"), according to the classification of Y. Wada and R. Hayakawa [11]. However, BPE and case “C” are not identical. The polymer with embedded charged regions is characterized by a dipole moment as whole, similar to BPE. In order to exhibit pyro- or piezo-response the polymer must be heterogeneous (a spatially inhomogeneous dielectric constant, or thermal expansion, or temperature). BPE does not require heterogeneity of this sort, though any barrier structure in a semiconductor reflects an inhomogeneity of its electronic properties. The origin of BPE, as well as that of the PNJ, is the contact potential difference between *p*- and *n*- semiconductors, or between metal and semiconductor for the Schottky contact. Thus, the origin and nature of BPE differs from those of other pyroelectric structures.

Moreover, the preparation technologies of polymer electrets and BPE-structures differ essentially. Polymer electrets require, e.g. charging by corona discharge [11, 12], while BPE-structures are formed with the creation of the barrier structure. Thus, BPE-structures can be considered as a new class of pyroelectrics.

Pyroelectricity and piezoelectricity are closely related phenomena - many pyroelectrics are also piezoelectrics. In particular, this relates also to polymer electrets [11, 12], and apparently also to barrier structures**.** Any barrier structure (PNJ, Schottky contact, heterojunction etc.) must also possess always a barrier piezoelectric effect (BPZE). Indeed, properties determining the width of the depleted region, hence, the magnitude of the dipole moment (e.g. doping concentrations, dielectric constant and energy spectrum), depend on deformation. Also, deformation is connected with generation or absorption of heat, and, hence leads to a pyroelectric contribution [13]. Thermodynamically, the inverse BPZE must exist as well. Observation of BPZE, in an organic semiconductor Schottky contact, has been reported recently [14].

One should emphasize that BPZE is an inherent property of a barrier structure in any material, and not only in quantum paraelectric semiconductors. Also, a strong temperature dependence of the dielectric constant, which is a crucial condition for the BPE, is not required for the BPZE. Thus, BPZE is even a more common phenomenon than BPE.

The aim of this work is an experimental demonstration of pyroelectric effect in barrier structures. The PbTe *p*-*n* junctions, studied here, are not the optimal system to reach the highest BPE, but they serve as a useful model system allowing exploring the main features of BPE.

## II. EXPERIMENTAL

### A. Sample preparation and characterization

The detailed account of preparation and characterization of the PbTe PNJs has been reported earlier [9,10]. For the sake of completeness we summarize here the main properties and results. The PNJ arrays have been prepared on a rectangular slice of a *p*-PbTe cut from a monocrystal ingot, grown by the Czochralski method. The acceptor concentration is $N_a = 10^{17}$ cm$^{-3}$. At 80 K the hole diffusion length was $\approx 30$ μm, and the mobility was $\approx 14500$ cm$^2$/Vsec. The PbTe absorption coefficient, at wave length of 10.5 $\mu$m, and carrier concentration of $10^{17}$ – $10^{18}$ cm$^{-3}$, is $\alpha$ = $10^2$ cm$^{-1}$ in the temperature interval of 20 - 200 K [15].

Two methods have been applied to create the $n$ -region: (1) thermodiffusion (TDJ) of In from $In_4Te_3$ gas phase, and (2) In or Zn ion-implantation (IJ). The junction was at a depth of $\approx 70$ $\mu$m under the surface in TDJ and a few $\mu$m in IJ.

The current-voltage (*I*-*V*) and capacitance-voltage (*C*-*V*) characteristics have been measured in the temperature region of 10 - 200 K. The *I*-*V* curves were fitted by the Shockley formula [16] with the ideality factor $n \approx 1.5 \div 2$, and saturation current density of $\approx$ $10^{-5}$ A/cm$^2$ at 80 K. The high-temperature activation energy of the zero-bias resistance ($R_0$) is $\approx 105$ - 110 meV. This is close to half of the PbTe-gap at 0 K, pointing to the recombination-generation mechanism of the current transport [17]. The low-temperature region with low activation energy indicates, probably, the presence of band-to-band tunneling via intermediate local states in the gap [9, 10].

The linearity of the $C^{-3}$-$V$ plots indicates that these PNJs are linearly graded [16]. The temperature dependence of the dielectric constant $\varepsilon(T)$, was derived from the temperature dependence of the capacitance. The dielectric constant is fitted by Barrett's formula for quantum paraelectrics [18]

$$\varepsilon = \frac{1.36\times 10^5}{36.14\times \coth\left(36.14/T\right)+49.15}, \qquad (1)$$

Thus, $\varepsilon$(300 K) $\approx 400$ and $\varepsilon$(100 K) $\approx 1600$.

The thermal properties of PbTe, used in the calculations, have been adopted from the literature [19-21].

B. The measurements.

The experimental set-up for measuring BPE is shown in Fig. 1(a). The pyroelectric signal (PES) was excited by heating the PNJ with a chopper modulated $CO_2$ laser with a wavelength of 10.6 μm. The photon energy is $\leq$ 120 meV, while the PbTe gap varies from 190 meV at 0 K to 320 meV at 300 K. Thus, a photovoltaic effect is excluded, and only the alternative heating-cooling process is possible.

The PES temporal evolution has been measured by a Tektronix Differential Preamplifier ADA400A and displayed on a Tektronix TDS 3054B scope. The PES signal with an applied bias, using a Keithley Sourcemeter 2410, has been measured by an EG&G Princeton Applied Research Lock-in Amplifier 5209 (LIA).

The maximal laser power applied was $\approx$ 1 $W/cm^2$. The signal was found to depend linearly on the laser power (Fig. 2). The chopping frequency (*f*) was varied from 4 Hz to 2000 Hz. The chopper slit was much wider than the width of the PNJ. Therefore, the dependence of the illuminated area on time was trapezoidal shaped, with short rise- and fall-times, and a long period of constant illumination, Fig. 1(b). The ratio of these times was about 1:10.

The excitation was applied in two modes. One was a periodic mode with equal durations of illumination and darkness, the period being shorter than both the temperature relaxation time ($t^*$) and the electronic relaxation time $\tau_e = RC$, of the diode. The periodic signal, stabilized after a few pulses, was then measured.

The other mode consisted of single pulse excitation (with a darkness period much longer than the duration of illumination), so that the temperature and *RC* relaxations have been completed before the arrival of the next pulse.

The measurements were carried out over a temperature interval of 12 - 130 K. The samples were placed in a closed cycle He-gas refrigerator cryostat, in a vacuum of about $10^{-7}$ torr. The temperature stability was 0.02 K above 100 K, and about 0.002 K below 100 K.

Fig. 2 shows typical examples of a periodic signal (a), and a single pulse signal (b) at zero bias. The magnitude of the signal is about $10 - 20\ \mu\text{V}$. It is clear that the measured response has all the characteristic features of a pyroelectric signal: (1) it is excited by the modulated light only; (2) the signal is excited by quanta with energy smaller than the PbTe gap; (3) and, perhaps, the most important mark - the signal has opposite signs at heating and cooling stages.

## III. THEORY

The theory of BPE in PNJ will be developed here, akin to the standard theory of pyroelectric effect for thin pyroelectric films [22-24], with specific attention paid to the junction depleted region [4, 16]. The BPE theory is based on the simultaneous solution of a pair of coupled equations: the thermal balance equation, and the Kirchoff's equation for quasi-stationary currents. The thermal balance is given by the equation

$$C_T \frac{dT}{dt} + G_T T = A(t) P, \qquad (2)$$

where $C_T$ is the total thermal capacity of the PNJ, $G_T$ is the heat transfer coefficient, $A(t)$ is the illuminated area of the sample and $P$ is the light intensity. Here $T = T_j - T_0$ is the difference between the actual junction temperature $T_j$ and the temperature $T_0$, as set by the temperature controller. The term $G_T T$ describes the Newton heat transfer.

The following comments are in order regarding the use of Eq. (2) in the present case. Equation (2) was formulated to describe the pyroelectric effect in thin films of standard pyroelectrics (wide-gap dielectrics). Therefore, it is based on the following assumptions: (1) It assumes that $\alpha L < 1$, where $\alpha$ is the absorption coefficient and $L$ is the width of the depletion region of the junction. Thus the energy, $A(t)P$, is distributed uniformly over the junction region; (2) Eq. (2) ignores thermal conductivity, and describes the loss of heat by the empirical Newton heat transfer coefficient, $G_T$ ; (3) respectively, the total thermal capacity is $C_T = V_S \rho c$ , where $V_S$ is the heated volume, $\rho$ is the mass density , and $c$ is the specific heat.

The condition of $\alpha L < 1$ is satisfied in this case [15]. However, points (2) and (3) deserve some reservation. The junction region (which plays here the role of the pyroelectric medium) is embedded in the PbTe crystal, and the effective heated volume $V_S$ varies during the heating-cooling cycles. Indeed, $V_S$ varies due to thermoconductivity as the heat is transferred into, or from, the adjoining PbTe regions during the heating or cooling stages. This problem would not arise, if the accurate thermoconductivity equation was applied, taking into account a $\nabla^2 T$ term instead of $G_T T$, and the corresponding boundary conditions. However, in the present case the geometry of the heated volume is too complex, thus it is advantageous to use Eq. (2), taking into consideration its limitations. These will be discussed later in this article.

The equation for the current is

$$\frac{d\left[C\left(V_t\right)V_t\right]}{dt} + \frac{U}{R\left(V_t\right)} = 0, \qquad (3)$$

where $V_t = V_0 - V_b + U$, $V_0$ is the PNJ built-in barrier potential, $V_b$ is the bias voltage ($V_b < 0$ for the reverse bias) [4,16], $U$ is the pyroelectric signal, $C(V_t)$ is the total capacitance of the PNJ and $R(V_t)$its resistance. Typical value of $V_0$ is a few tens of mV, and of $V_b$ up to a few hundreds of mV [10]. Since $U << V_0$ (Fig. 2), one can expand all terms in Eq. (3) into a power series in $U$, keeping linear terms only. Then, designating $V = V_0 - V_b$, we have

$$C(V)\frac{dU}{dt}+\frac{U}{R(V)}=\Lambda\frac{dT}{dt} \tag{4}$$

$$\Lambda=-V\frac{dC(V)}{dT}. \tag{4.1}$$

The left-hand part of Eq. (4) has dimensions of electrical current, thus Λ may be interpreted as the "current" pyroelectric coefficient [22-24], and $C(V)$, being the total capacitance of the PNJ, is expressed in C/K. This differs from the standard definition of the pyroelectric constant, having dimensions $C/cm^2$ K. In our case the total capacity, and not the capacitance per square area, is a property of sample. It is thus obvious, that standard specific pyroelectric coefficient is

$$\lambda = \Lambda/A_e \tag{4.2}$$

where $A_e$ is the area of the *p*-*n* junction. Eq. (4) has a clear physical meaning. This is a standard equation for an alternating current source, with intensity determined by the temperature modulation of the built-in dipole moment. When the barrier height is zero, $V_0 = 0$, the bias $V_b$ is zero as well (as it is impossible to apply the bias when the barrier is absent), thus the source vanishes, i.e. $V = 0$. Thus, when the barrier vanishes, so does Λ (Eq. 4.1).

The left-hand part of Eq. (4) consists of two currents

$$I_1=C(V)\frac{dU}{dt}\ \text{and}\ I_2=\frac{U}{R(V)}. \tag{5}$$

$I_1$ is the displacement current arising due to the change of the PNJ dipole moment, stimulated by the change of temperature; $I_2$ is the conduction current, screening the non-equilibrium polarization of the PNJ. The ratio of these currents is

$$\frac{I_2}{I_1}=\frac{1}{RC\left(d\ln U/dt\right)} \tag{6}$$

Eq. (4) can be rewritten in the form

$$\frac{dU}{dt}+\frac{U}{R(V)C(V)}=\frac{\Lambda}{C(V)}\frac{dT}{dt} \tag{7.1}$$

We define now

$$\Lambda_U=-V\frac{d\ln C(V)}{dT}=\frac{\Lambda}{C(V)};\quad \lambda_U=\frac{\Lambda_U}{A_e} \tag{7.2}$$

In the following, we use this $\Lambda_U$, since all measurements have been carried out in the constant current mode. $\Lambda_U$ will be termed the "voltage pyroelectric coefficient". According to Eq. (4.1), the value of $\Lambda$ varies with doping (via $V_0$), with the dopant concentration gradient (via $C$), with $\varepsilon(T)$ (via $C$), as well as with the bias and the temperature [1, 3, 4, 9, 10].

All values on the right-hand side of Eq. (7.1) are known from the diode characterization data. The pyrocoefficients $\lambda$ and $\lambda_U$ for the PbTe PNJ are presented in Figs. 3(a) and 3(b). The values are negative, corresponding to $d\varepsilon/dT<0$, [see Eq. (1)].

To summarize, the pair of coupled equations (2) and (4) govern the BPE in PNJ. The general solutions of these equations are

$$T(t)=\frac{P}{C_T}\exp\left(-\frac{t}{t^*}\right)\int_0^t dt'\exp\left(\frac{t'}{t^*}\right)A(t')+T(0);\ t^*=\frac{C_T}{G_T} \tag{8}$$

$$U(t)=\Lambda_U\exp\left(-\frac{t}{\tau_e}\right)\int_0^t dt'\exp\left(\frac{t'}{\tau_e}\right)\frac{dT(t')}{dt'}+U(0);\quad \tau_e=RC \tag{9}$$

The source in Eq. (9) is $dT/dt$. Using Eq. (8) this can be expressed as

$$\frac{dT}{dt}=\frac{P}{C_T}\left\{A(t)-\frac{1}{t^*}\exp\left(-\frac{t}{t^*}\right)\int_0^t dt'\exp\left(\frac{t'}{t^*}\right)A(t')\right\}. \tag{10}$$

The source consists of two contributions. The positive first term is the rate of the energy generation, while the negative second term is the rate of heat dissipation.

Eq. (4) can be considered also independently. It can be applied to reconstruct the change of temperature within the PNJ for an arbitrary, BPE generating, source. Indeed, Eq. (4) can be rewritten as follows:

$$\frac{dT}{dt}=\frac{C(V)}{\Lambda}\frac{dU}{dt}+\frac{U}{R(V)\Lambda} \tag{11}$$

The right-hand part of Eq. (11) is a known function of time at given temperature, and bias. In fact, all the terms at the right-hand side are known from the experimentally determined kinetics of the signal, and from the diode characterization. Thus, using Eq. (11), $dT/dt$ is defined and further integration provides $\Delta T(T)$- the kinetics of temperature evolution within the junction region.

To find the number of independent parameters entering into Eqs (2) and (7), we introduce the following dimensionless functions and arguments:

$$\frac{T}{T^*}=z;\ \frac{t}{t^*}=y;\ \frac{U}{U^*}=w;\ T^*=\frac{2\pi HrfPC_T}{G_T^2};\ t^*=\frac{C_T}{G_T};\ U^*=\Lambda_U T^*, \tag{12}$$

where $H$ is the height of chopper slit along its radius, $r$ is the distance of the slit from the chopper axis, and $f$ is the rotational frequency of the chopper [$2\pi Hrf$ is the area swept out by the slit edge per unit time (Fig. A1.1, Appendix A1)]. Expressing the signal in units of $U^*$, justifies calling $\Lambda_U$ "the voltage pyrocoefficient". Let us introduce also the following notation

$$p=\frac{t^*}{\tau_e} \tag{13}$$

Now, Eqs. (2) and (7.1) assume the form (Appendix A1)

$$\frac{dz}{dy}+z=S(y); \tag{14}$$

$$S(y)=\begin{cases} y; & 0<y<y_1 \\ y_1; & y_1<y<y_1+y_2; \\ 2y_1+y_2-y; & y_1+y_2<y<2y_1+y_2; \\ 0; & 2y_1+y_2<y<2(y_1+y_2) \end{cases} \tag{15}$$

$$\frac{dw}{dy}+p\,w=\frac{dz}{dy} \tag{16}$$

The dimensionless time segments $y_1$ and $y_2$ correspond to the times $t_1$ and $t_2$ in Fig. 1(b).

Eq. (14) shows that the kinetics of thermal process is characterized by the thermal relaxation time $t^*$. The pyroelectric signal, according to Eqs. (15) and (13), is governed by two relaxation times: $t^*$ and the electronic relaxation time $\tau_e = RC$. Eq. (14) and Eq. (15) are solved in accordance with the experimental modes, namely: single laser pulse excitation and periodic excitation.

*1. The single pulse excitation.*

The solution (Appendices A1 and A2) must satisfy the boundary conditions and the continuity conditions at the ends of the time intervals ($y_1$, $y_1+y_2$, $2y_1+y_2$). This solution describes completely the relaxations connected with both characteristic times. The temperature evolution is then given by the function $z(y)$ (derived in Appendix 1),

$$\begin{aligned} z(y)=&\left(y-1+e^{-y}\right)\Theta\left(y_1-y\right)+\\ &+\Theta\left[y-\left(y_1+\delta\right)\right]\left[y_1-\left(e^{y_1}-1\right)e^{-y}\right]\Theta\left[\left(y_1+y_2\right)-y\right]+\\ &+\Theta\left[y-\left(y_1+y_2+\delta\right)\right]\left[\left(2y_1+y_2=1\right)-y+\left(1-e^{y_1+y_2}-e^{y_1}\right)e^{-y}\right]\Theta\left[2y_1+y_2-y\right]+\\ &+\Theta\left[y-\left(2y_1+y_2+\delta\right)\right]\left[\left(e^{y_1+y_2}-1\right)\left(e^{y_1}-1\right)e^{-y}\right], \end{aligned} \tag{17}$$

where

$$\Theta(x)=\begin{cases} 0, & x<0 \\ 1, & x\ge 0 \end{cases}, \tag{17.1}$$

and $\delta \ll y_1, y_2$. (The introduction of $\delta$ is necessary due to the $\geq$ sign in Eq. (17.1), note in Ref. [25]**)**.

The solution of the excited voltage signal is

$$\begin{aligned} w = {} & w_1\Theta\left(y_1 - y\right) + \\ & + w_2\Theta\left[y - \left(y_1 + \delta\right)\right]\Theta\left(y_1 + y_2 - y\right) + \\ & + w_3\Theta\left[y - \left(y_1 + y_2 + \delta\right)\right]\Theta\left(2y_1 + y_2 - y\right) + \\ & + w_4\Theta\left[y - \left(2y_1 + y_2 + \delta\right)\right] \end{aligned} \tag{18}$$

and the coefficients $w_1$, $w_2$, $w_3$ and $w_4$ are defined in Appendix A2.

2. *The periodic excitation.*

The second case is a stable periodic excitation generated by periodic laser heating. The solution is constructed by expanding both sides of Eqs. (14) and (16) into Fourier series (Appendix A3). The temperature and voltage signals are, correspondingly

$$z\left(y\right) = \frac{y_1}{2} + \sum_{n=1}^{\infty} \frac{c_n - \frac{n\pi}{l} d_n}{1 + \left(\frac{n\pi}{l}\right)^2} \cos\frac{n\pi y}{l} + \frac{\frac{n\pi}{l} c_n + d_n}{1 + \left(\frac{n\pi}{l}\right)^2} \sin\frac{n\pi y}{l};\ l = y_1 + y_2; \tag{19}$$

$$\begin{aligned} w\left(y\right) = {} & \sum_{n=1}^{\infty} \frac{\frac{n\pi}{l}}{\left[p^2 + \left(\frac{n\pi}{l}\right)^2\right]\left[1 + \left(\frac{n\pi}{l}\right)^2\right]} \times \\ & \left[\left\{\frac{n\pi}{l}\left(p+1\right)c_n + \left[p - \left(\frac{n\pi}{l}\right)^2\right]\right\}\cos\frac{n\pi y}{l} + \left\{-\left[p - \left(\frac{n\pi}{l}\right)^2\right]c_n + \frac{n\pi}{l}\left(p+1\right)d_n\right\}\sin\frac{n\pi y}{l}\right] \end{aligned} \tag{20}$$

and the coefficients $c_n$ and $d_n$ are defined in Appendix A3.

To obtain final smooth solutions of (19) and (20) about 15 terms of the series had to be summed.

## IV. RESULTS AND DISCUSSION

We want to show that the theory developed here describes, adequately, quantitatively the experimental data and allows determining the fitting parameters $C_T$, $G_T$ and $p$.

The experimental results are presented in Figs. 4 - 9.

Figs. 4(a1) – 4(a5) shows the kinetics of the periodic signal at $T = const$, at different chopper frequencies and at zero bias ($V_b = 0$). For comparison, the corresponding calculated curves, Figs. 4(b1) – 4(b5) are also depicted in this figure. Figs. 5(a1) – 5(a5) shows the time evolution of the signal at $f = const$, at different temperatures and at zero bias. Figs. 5(b1) – 5(b5) are the corresponding calculated curves. Fig. 6 depicts the dependence of the amplitude of the periodic signal on frequency and temperature, at zero bias. Fig. 7 shows the dependence of the amplitude of the periodic signal on bias.

The main features of these dependencies are as follows: (1) The magnitude of the signal is ~ $10 - 20\,\mu\text{V}$ . The signal exhibits the typical pyroelectric response shape: a rather sharp increase at the onset of the illumination, then the plateau or a slowly decreasing phase, followed by a sharp drop accompanied by a change of the sign at the end of the light pulse; this is then followed by the negative replica of the light-on period. (2) The shape of the signal varies rather markedly with frequency. However, the amplitude of the signal is, practically, frequency independent. (3) The amplitude of the signal changes non-monotonically with temperature, dropping at higher temperatures. (4) The amplitude of the signal increases with increasing reverse bias.

A rather cumbersome, but straightforward, procedure was performed to find the parameters of the dimensionless theoretical signal $w(y, y_1, y_2, p)$ (Eq. (20), that give the best fit to the measured signal $U(t)$, as shown in Figs. 4, 5. The ratio of the amplitudes of the measured signal ($U_m$) and of the calculated signal ($w_m$) gives the normalizing voltage $U^*$ = $U_m/w_m$. It is seen that the calculations reproduce satisfactorily well the features of the experimentally observed signal.

Thus, the fitting procedure leads to two equalities with left-hand parts determined experimentally

$$t^*(T, f) = \frac{C_T}{G_T} = \frac{V_S c\rho}{G_T} \quad \text{and}$$
$$U^*(T, f, V_b) = \Lambda_U T^* = \frac{2\pi HrfPC_T}{G_T^2} = \Lambda_U \frac{2\pi HrfPt^*}{G_T} \tag{21}$$

This allows determining $C_T$ and $G_T$, or $G_T$ and $V_S$.

The amplitude $U_m$ drops at higher temperatures, as $U_m \propto \Lambda_U \propto V_0$ [Eqs. (9), (7.2)] and the built-in barrier $V_0$ decreases at higher temperatures.

The dependence of $U_m$ on $V_b$ is the result of the corresponding dependence on $V_b$ of $\lambda_U$ [Fig. 3(b), Eq. (7.2)] and $\tau_e$. The signal disappears in the regime of positive bias, evidently due to the disappearance of the barrier and the corresponding vanishing of the dipole moment.

Fig. 8 present the kinetics of the single pulse response and the result of fitting at $T$ = 12.5 K, and different duration of illumination $(\Delta t)$. The agreement between the experimental and calculated functions is clearly evident.

The magnitude of the signal depends weakly on $\Delta t$, (the analog of the periodic signal dependence on frequency). The signal decreases with increase of temperature, as also observed for the periodic signal. This is due to disappearance of PNJ at high temperature.

In Section III**,** the method of deriving the temperature kinetics $\Delta T(t)$ inside the junction has been described [see Eq. (11)]. The successive steps of this procedure and the results, for the periodic signal, are presented in Figs. 9-11.

Fig. 9(a) shows the measured signal $U(t)$; Fig. 9(b) shows $dU/dt$ calculated from the data in Fig. 9(a); Fig. 10(a) shows the $dT/dt$ calculated from the data; Fig. 10(b) shows the $\Delta T\left(t\right)$ obtained by integration of the data in Fig. 10(a). All four traces are synchronized. Figs. 9-10 present data taken at $T$ = 25.5 K, and at different frequencies. Data taken at $f$ = 40 Hz and at different temperatures, lead to similar results.

The following conclusions are drawn from these graphs.

(1) The amplitude of $\Delta T$ is ~ 10 mK at low temperature and exceeds ~1 K at high temperatures (Fig. 11). We interpret this, as result of decreasing the thermodiffusivity at high temperature [19-21]. Therefore, the heat transfer between the junction region and the adjoining media, during the characteristic period $\Delta t$ decreases. In other words, the effective heated volume $V_S$ decreases.

(2) For the same reason the value of $dT/dt$ increases also with temperature. The amplitude of $dT/dt$ increases with frequency.

(3) These graphs show clearly the limitations of the thermobalance Eq. (2). Indeed, it is evident that Eq. (2) can give only a rise of temperature during the illumination period. However, Fig. 10 shows that with an increase of $\Delta t$, starting from $\approx 4.6$ milliseconds, the temperature decreases even during the illumination. This is due to heat transfer from the PNJ into the adjoining PbTe bulk. This is clearly apparent at longer $\Delta t$ and lower temperature, when the time is long enough, and the thermal diffusion is high. This thermoconductivity process is not accounted for in the simplified model as expressed by Eq. (2).

The total pyroelectric current $I$ consists of the displacement current $I_1$ and the screening conduction current $I_2$ [Eqs. (5) and (6)]. Referring to Figs. 12 and 13, one can analyze their experimental behavior, as follows: (1) The current amplitude increases from 0.1 nA at low temperature to 1.5 nA at high temperature; (2) $I_2 << I_1$ at low temperature, and $I_2 >> I_1$ at high temperature. This is due to the exponential decrease of the PNJ resistance when increasing the temperature;[10] (3) It is also apparent that the screening current $I_2$ lags behind the displacement current $I_1$. This expresses the physical fact that it is the change of polarization (the displacement current) which initiates the screening conduction current.

Figs. 13 shows the time evolution of the junction temperature $\Delta T(t)$ and the correlation of the currents $I_1$ and $I_2$ for single pulse excitation. The picture is similar to that of the periodic signal.

Fig. 14(a) presents $t^*(T, f)$. The value of $t^*$ increases with $\Delta t$ dependence as conditioned by an increase of $V_S$ during warming. At low temperature $t^*$ increases because of the increase of thermodiffusivity [19-21]. The heat transfer coefficient $G_T(T)$ in Fig. 14(b) follows the variation of the thermoconductivity.

Fig. 15 displays $V_S(T, f)$. The relation $V_S \propto f^{1/2}$ is typical for a one-dimensional case, reflecting the fact that the thickness of PNJ is the smallest dimension.

Fig. 16 presents the parameter $p(T, f) = t^*/\tau_e$. It is clear that the behavior is determined by the strong temperature dependence of the diode resistance.

Similar behavior of these properties has been obtained in the single pulse case.

## V. CONCLUSION

We have demonstrated here the existence of a pyroelectric response of a barrier structure in the non-polar semiconductor. Such structures present a new class of pyroelectrics, and are fundamentally different from the conventional PE systems.

We have observed a similar effect in the Schottky contact of an In-*p*PbTe.

As semiconductors, these systems allow new ways of controlling the PE-response, e. g. by doping or varying the temperature or bias voltage, by controlling the dipole moment of the junction by photoactive illumination etc. We have shown also that BPE depends on the diode resistance. Hence, magnetic field and deformation will affect the PE response.

The BPE in the PbTe *p-n* junctions is markedly smaller, than in the best commercial pyroelectrics (PZT). However, BPE will be much higher in the $ABO_3$ *p-n* junctions. The success of laser MBE preparation technology of such film systems has been reported [26]. An increase of the PE response may also be attained by creation of barrier superlattices.

The theoretical analysis of the BPE data has been successfully demonstrated, as well as the reconstruction of the temperature kinetics within the *p-n* junction.

## ACKNOWLEDGMENTS

The authors wish to express their gratitude to Prof. A. Shaulov from the Department of Physics at Bar-Ilan University, for his enlightening comments and willingness to share his experience. One of the authors (R. K.) acknowledges Bar-Ilan University President’s Fund scholarship. This work is part in the fulfillment of the requirements toward his PhD.

## APPENDIX A1:
## TEMPERATURE KINETICS FOR LINEARLY INCREASING ILLUMINATED AREA AND SINGLE PULSE EXCITATION - THE RELAXATION SOLUTION.

The sample area is a square of 0.8 mm × 0.8 mm. The beam crossection is a circle with a diameter *L* of 1.2 mm. The mean radius of the chopper window is *r* = 41.5 mm. The mean length of the window is *D* = 21.5 mm (see Fig. A1.1). The height of the window is $H \approx 15$ mm, but since the beam diameter and *H* are larger than the height of the sample, we used in calculations the effective value of *H* = 0.8 mm.

In the present case *D*>>*L*. The linear velocity of the window edge is $v = 2\pi f r$, where *f* is the rotation frequency and *r* the mean radius of the window arc. Let *t*=0 be the moment when the chopper window starts to overlap spot of the laser beam. The beam opens fully at $t_1 = L/v$. The light starts closing down at $t = D/v$. The full illumination interval is $t_2 = (D-L)/v$. The light gets totally closed at $t_2 + t_1 = (D+L)/v$. It will stay closed until $2t_1 + 2t_2 = 2D/v$.

The thermo-balance equation is

$$C_T \frac{dT}{dt} + G_T T = A(t)P; \qquad \text{(A1.1)}$$

where

$$A(t) = \begin{cases} b\,t, \quad 0 \le t < t_1; \\ A(t_1) = b\,t_1 = A_m; \quad t_1 < t < t_1 + t_2; \text{ light-on} \\ A_m - b\left[t - \left(t_1 + t_2\right)\right]; \quad t_1 + t_2 < t < 2t_1 + t_2; \\ 0; \quad 2t_1 + t_2 < t < 2\left(t_1 + t_2\right); \text{ light-off} \end{cases} \qquad \text{(A1.1.1)}$$

$$b = Hv = 2\pi Hrf; A_m = H\,L;$$

$$t_1 = \frac{L}{2\pi rf}; \quad t_2 = \frac{D-L}{2\pi rf}.$$

*P* is the power of illumination; $C_T$ is thermocapacity of the sample; $G_T$ is the heat-loss factor.

Let us define new variables:

$$\frac{T}{T^*} = z; \quad \frac{t}{t^*} = y; \qquad \text{(A1.1.2)}$$

where

$$T^* = \frac{2\pi HrPfC_T}{G_T{}^2} \text{ and } t^* = \frac{C_T}{G_T}, \quad \text{(A1.1.3)}$$

then Eq. (A1.1) can be rewritten as follows:

$$\frac{dz}{dy} + z = \frac{P}{T^* G_T} A; \; z(0) = 0 \quad \text{(A1.2)}$$

The dimensionless equation (A1.2) is then applied to each of the four temporal regions, see Fig. 1(b), as shown below in Eqs. (A1.3). The initial condition for the solution in a time domain is given by the end value of the function *z*(*y*) in the preceding time interval. The solution is related to the single pulse case, when the next light pulse occurs after a very long dark period. Then, the fourth part region (see below) is $2y_1+y_2 < y < \infty$. The dark period is very long, so that the relaxation of temperature terminates completely.

$$\begin{aligned}
&1)\; 0 < y < y_1;\; z_1(0) = 0; \\
&\quad \frac{dz_1}{dy} + z_1 = y \\
&2)\; y_1 < y < y_1 + y_2;\; z_2(y_1) = z_1(y_1); \\
&\quad \frac{dz_2}{dy} + z_2 = y_1 \\
&3)\; y_1 + y_2 < y < 2y_1 + y_2;\; z_3(y_1 + y_2) = z_2(y_1 + y_2); \\
&\quad \frac{dz_3}{dy} + z_3 = 2y_1 + y_2 - y \\
&4)\; 2y_1 + y_2 < y < 2(y_1 + y_2);\; z_4(2y_1 + y_2) = z_3(2y_1 + y_2) \\
&\quad \frac{dz_4}{dy} + z_4 = 0
\end{aligned} \quad \text{(A1.3)}$$

The solutions of Eqs. (A1.3) are:

1) $0 < y < y_1$;

$$z_1 = y - 1 + e^{-y};$$

$$z_1(y_1) = y_1 - 1 + e^{-y_1};\ \frac{dz_1}{dy} = 1 - e^{-y} > 0;\ \left.\frac{dz_1}{dy}\right|_0 = 0;\ \left.\frac{dz_1}{dy}\right|_{y_1} = 1 - e^{-y_1} \tag{A1.4}$$

As $z_1(0) = 0$, and $dz_1/dy > 0$, so $z_1(y) > 0$, and in particular, $z_1(y_1) > 0$.

2) $y_1 < y < y_1 + y_2$;

$$z_2 = y_1 - \left(e^{y_1} - 1\right)e^{-y};$$

$$\frac{dz_2}{dy} = \left(e^{y_1} - 1\right)\cdot e^{-y} > 0;$$

$$\left.\frac{dz_2}{dy}\right|_{y_1} = 1 - e^{-y_1};\ z_2(y_1 + y_2) = y_1 - \left(1 - e^{-y_1}\right)e^{-y_2} \tag{A1.5}$$

As $z_2(y_1) > 0$, and $dz_2/dy > 0$, so $z_2(y) > 0$.

$$\left.\frac{dz_2}{dy}\right|_{y_1+y_2} = \left(1 - e^{-y_1}\right)\cdot e^{-y_2}$$

From Eqs. (A1.4) and (A1.5), *dz/dy* increases in this region and *dz/dy* is continuous at $y_1$.

3) $y_1 + y_2 < y < 2y_1 + y_2$;

$$z_3 = (2y_1 + y_2 + 1) - y + \left(1 - e^{y_1+y_2} - e^{y_1}\right)e^{-y};$$

$$z_3(2y_1 + y_2) = \left(1 - e^{-y_1}\right)\left(1 - e^{-y_1-y_2}\right);$$

$$\frac{dz_3}{dy} = -1 - \left(e^{-y_1} - e^{y_2} - 1\right)\cdot e^{y_1-y};\ \left.\frac{dz_3}{dy}\right|_{y_1+y_2} = \left(1 - e^{-y_1}\right)e^{-y_2} \tag{A1.6}$$

$$\left.\frac{dz_3}{dy}\right|_{y_0} = 0;\ \ y_0 = \ln\left(e^{y_1} - 1 + e^{y_1+y_2}\right).$$

4) $2y_1 + y_2 < y < 2(y_1 + y_2)$;

$$\begin{aligned}
&z_4 = \left(e^{y_1+y_2} - 1\right)\left(e^{y_1} - 1\right)e^{-y};\\
&z_4\big|_{2y_1+y_2} = z_3(2y_1 + y_2) = \left(1 - e^{-y_1}\right)\left(1 - e^{-y_1-y_2}\right);\\
&\frac{dz_4}{dy} = -\left(e^{y_1+y_2} - 1\right)\left(e^{y_1} - 1\right)e^{-y} < 0\\
&z_4\left[2(y_1 + y_2)\right] = \left(e^{y_1+y_2} - 1\right)\left(e^{y_1} - 1\right)e^{-2(y_1+y_2)}
\end{aligned} \tag{A1.7}$$

In this region, $z_4$ decreases as exp*(-y)*. If we reconstruct $\Delta T(t)$ in this region, then from $\ln\Delta T(t)$ vs $t$ one calculates $t^*$.

Using the above listed physical data, the following values of $t_1$ and $t_2$ were calculated:

| | | |
|---|---|---|
| For $f$ = 1 Hz | $t_1 = 4.6 \times 10^{-3}$ s; | $t_2 = 8.1 \times 10^{-2}$ s |
| For $f$ = 20 Hz | $t_1 = 2.3 \times 10^{-4}$ s; | $t_2 = 4.05 \times 10^{-3}$ s |
| For $f$ = 800 Hz | $t_1 = 5.75 \times 10^{-6}$ s | $t_2 = 1.01 \times 10^{-4}$ s |

## APPENDIX A2:
## THE VOLTAGE KINETICS FOR THE SINGLE PULSE EXCITATION.

Equation (16)

$$\frac{dw}{dy} + p\,w = \frac{dz}{dy}$$

is solved in the four time intervals, applying the initial and continuity conditions.

$$\begin{aligned}
&0 < y < y_1;\ w_1(0) = 0; \\
&\quad \frac{dw_1}{dy} + p\,w_1 = 1 - e^{-y} \\
&y_1 < y < y_1 + y_2;\ w_2(y_1) = w_1(y_1) \\
&\quad \frac{dw_2}{dy} + p\,w_2 = \left(e^{y_1} - 1\right)e^{-y} \\
&y_1 + y_2 < y < 2y_1 + y_2; \\
&\quad \frac{dw_3}{dy} + p\,w_3 = -1 - \left(1 - e^{y_1+y_2} - e^{y_1}\right)e^{-y} \\
&2y_1 + y_2 < y < \infty \\
&\quad \frac{dw_4}{dy} + p\,w_4 = -\left(e^{y_1+y_2} - 1\right)\left(e^{y_1} - 1\right)e^{-y}
\end{aligned} \qquad \text{(A2.1)}$$

The corresponding solutions are:

$$\begin{aligned}
w_1 &= \frac{1}{p} - \frac{1}{p-1}e^{-y} + \frac{1}{p(p-1)}e^{-py} \\
w_2 &= \frac{e^{y_1} - 1}{p-1}e^{-y} + \frac{1 - e^{py_1}}{p(p-1)}e^{-py} \\
w_3 &= -\frac{1}{p} - \frac{\left(1 - e^{y_1+y_2} - e^{y_1}\right)e^{-y}}{p-1} + \frac{1 - e^{py_1} - e^{p(y_1+y_2)}}{p(p-1)}e^{-py} \\
w_4 &= -\frac{\left(e^{y_1+y_2} - 1\right)\left(e^{y_1} - 1\right)}{p-1}e^{-y} + \frac{\left[e^{p(y_1+y_2)} - 1\right]\left(e^{py_1} - 1\right)}{p(p-1)}e^{-py}
\end{aligned} \qquad \text{(A2.2)}$$

Let us denote

$$\Theta(x) = \begin{cases} 0, & x < 0 \\ 1, & x \geq 0 \end{cases}$$

Then:

$$\begin{aligned} w(y) = {} & w_1\Theta(y_1 - y) + \\ & + w_2\Theta[y - (y_1 + \delta)]\Theta(y_1 + y_2 - y) + \\ & + w_3\Theta[y - (y_1 + y_2 + \delta)]\Theta(2y_1 + y_2 - y) + \\ & + w_4\Theta[y - (2y_1 + y_2 + \delta)] \end{aligned} \tag{A2.3}$$

where $\delta << y_1$. Because of the $\geq$ sign in the definition of $\Theta(y)$, insertion of $\delta$ prevents the addition of the contribution of two adjoining parts at the common point *y*.

# APPENDIX A3:
# SOLUTION OF THE TEMPERATURE AND VOLTAGE TIME DEPENDENCE –
# THE PERIODIC EXCITATION CASE.

The temperature kinetics is given by Eq. (14):

$$\frac{dz}{dy} + z = S(y). \tag{A3.1}$$

The equation is solved in the four time intervals as follows:

$$S(y) = \begin{cases} y; & 0 < y < y_1 \\ y_1; & y_1 < y < y_1 + y_2; \\ 2y_1 + y_2 - y; & y_1 + y_2 < y < 2y_1 + y_2; \\ 0; & 2y_1 + y_2 < y < 2(y_1 + y_2) \end{cases} \tag{A3.1.1}$$

To find the periodic solution, all terms of Eq. (A3.1) are expanded into Fourier series.

$$z = \frac{a_0}{2} + \sum_{n=1}^{\infty}\left(a_n \cos\frac{n\pi y}{l} + b_n \sin\frac{n\pi y}{l}\right); \; l = y_1 + y_2 \tag{A3.2}$$

$$a_n = \frac{1}{l}\int_0^{2l} z(y)\cos\frac{n\pi y}{l}dy; \; n = 0,1,2,..... \tag{A3.2.1}$$

$$b_n = \frac{1}{l}\int_0^{2l} z(y)\sin\frac{n\pi y}{l}dy; \; n = 1,2,..... \tag{A3.2.2}$$

$$a_0 = y_1 \tag{A3.2.3}$$

$$\frac{dz}{dy} = \sum_{n=1}^{\infty}\frac{n\pi}{l}\left(-a_n \sin\frac{n\pi y}{l} + b_n \cos\frac{n\pi y}{l}\right) \tag{A3.3}$$

$$S = \frac{c_0}{2} + \sum_{n=1}^{\infty}\left(c_n \cos\frac{n\pi y}{l} + d_n \sin\frac{n\pi y}{l}\right), \tag{A3.4}$$

$$c_n = \frac{1}{l}\int_0^{2l} S(y)\cos\frac{n\pi y}{l}dy; \; n = 0,1,2,... \tag{A3.4.1}$$

$$d_n = \frac{1}{l}\int_0^{2l} S(y)\sin\frac{n\pi y}{l}dy; \; n = 1,2,... \tag{A3.4.2}$$

$$c_0 = y_1 \tag{A3.4.3}$$

Evaluating the integrals in (A3.4.1) and in (A3.4.2) yields:

$$c_n = \frac{1}{y_1 + y_2} \left\{ \begin{array}{l} \left(\frac{l}{n\pi}\right)\left[ y_1 \sin\frac{n\pi y_1}{l} + \frac{l}{n\pi}\left(\cos\frac{n\pi y_1}{l} - 1\right)\right] + \\ + \frac{2y_1 l}{n\pi}\cos\frac{n\pi(2y_1+y_2)}{2l}\sin\frac{n\pi y_2}{2l} + \\ + \frac{2(2y_1+y_2)l}{n\pi}\sin\frac{n\pi y_1}{2l}\cos\frac{n\pi(3y_1+2y_2)}{2l} + \\ - \frac{(2y_1+y_2)l}{n\pi}\sin\frac{n\pi(2y_1+y_2)}{l} + \\ + \frac{(y_1+y_2)l}{n\pi}\sin\frac{n\pi(y_1+y_2)}{l} + \\ +2\left(\frac{l}{n\pi}\right)^2 \sin\frac{n\pi(3y_1+2y_2)}{2l}\sin\frac{n\pi y_1}{2l} \end{array} \right\} \qquad \text{(A3.5.1)}$$

$$.\ d_n = \frac{1}{y_1 + y_2} \left\{ \begin{array}{l} \left(\frac{l}{n\pi}\right)\left[ -y_1 \cos\frac{n\pi y_1}{l} + \frac{l}{n\pi}\sin\frac{n\pi y_1}{l}\right] + \\ + \frac{2y_1 l}{n\pi}\sin\frac{n\pi(2y_1+y_2)}{2l}\sin\frac{n\pi y_2}{2l} + \\ + \frac{2(2y_1+y_2)l}{n\pi}\sin\frac{n\pi y_1}{2l}\sin\frac{n\pi(3y_1+2y_2)}{2l} + \\ + \frac{(2y_1+y_2)l}{n\pi}\cos\frac{n\pi(2y_1+y_2)}{l} + \\ - \frac{(y_1+y_2)l}{n\pi}\cos\frac{n\pi(y_1+y_2)}{l} + \\ -2\left(\frac{l}{n\pi}\right)^2 \sin\frac{n\pi y_1}{2l}\cos\frac{n\pi(3y_1+2y_2)}{2l} \end{array} \right\} \qquad \text{(A3.5.2)}$$

Inserting the above expressions into Eq. (A3.1) one obtains the following relations:

$$a_n = \frac{c_n - \frac{n\pi}{l} d_n}{1 + \left(\frac{n\pi}{l}\right)^2}; \qquad \text{(A3.6.1)}$$

$$b_n = \frac{\frac{n\pi}{l} c_n + d_n}{1 + \left(\frac{n\pi}{l}\right)^2} \qquad \text{(A3.6.2)}$$

Then,

$$z=\frac{y_1}{2}+\sum_{n=1}^{\infty}\frac{c_n-\frac{n\pi}{l}d_n}{1+\left(\frac{n\pi}{l}\right)^2}\cos\frac{n\pi y}{l}+\frac{\frac{n\pi}{l}c_n+d_n}{1+\left(\frac{n\pi}{l}\right)^2}\sin\frac{n\pi y}{l};\ l=y_1+y_2 \qquad \text{(A3.7)}$$

$$\frac{dz}{dy}=\sum_{n=1}^{\infty}\frac{n\pi}{l}\left[\frac{\frac{n\pi}{l}c_n+d_n}{1+\left(\frac{n\pi}{l}\right)^2}\cos\frac{n\pi y}{l}-\frac{c_n-\frac{n\pi}{l}d_n}{1+\left(\frac{n\pi}{l}\right)^2}\sin\frac{n\pi y}{l}\right] \qquad \text{(A3.8)}$$

Expanding the function $w(y)$ into a Fourier series:

$$w=\frac{A_0}{2}+\sum_{n=1}^{\infty}A_n\cos\frac{n\pi y}{l}+B_n\sin\frac{n\pi y}{l};l=y_1+y_2 \qquad \text{(A3.9.1)}$$

$$\frac{dw}{dy}=\sum_{n=1}^{\infty}\frac{n\pi}{l}\left[B_n\cos\frac{n\pi y}{l}-A_n\sin\frac{n\pi y}{l}\right] \qquad \text{(A3.9.2)}$$

and using the expression of Eq. (A3.8) in Eq. (16) yields:

$$\frac{dw}{dy}+p\,w=\frac{dz}{dy}$$

and one obtains the following relations:

$$A_n=\frac{n\pi}{l}\frac{\frac{n\pi}{l}(p+1)c_n+\left[p-\left(\frac{n\pi}{l}\right)^2\right]d_n}{\left[p^2+\left(\frac{n\pi}{l}\right)^2\right]\left[1+\left(\frac{n\pi}{l}\right)^2\right]} \qquad \text{(A3.9.3)}$$

$$B_n=\frac{n\pi}{l}\frac{-\left[p-\left(\frac{n\pi}{l}\right)^2\right]c_n+\frac{n\pi}{l}(p+1)d_n}{\left[p^2+\left(\frac{n\pi}{l}\right)^2\right]\left[1+\left(\frac{n\pi}{l}\right)^2\right]} \qquad \text{(A3.9.4)}$$

resulting in :

$$w(y)=\sum_{n=1}^{\infty}\frac{\frac{n\pi}{l}}{\left[p^2+\left(\frac{n\pi}{l}\right)^2\right]\left[1+\left(\frac{n\pi}{l}\right)^2\right]}\times$$

$$\times\left(\begin{array}{l}\left\{\frac{n\pi}{l}\left(p+1\right)c_n+\left[p-\left(\frac{n\pi}{l}\right)^2\right]\right\}\cos\frac{n\pi y}{l}+\\+\left\{-\left[p-\left(\frac{n\pi}{l}\right)^2\right]c_n+\frac{n\pi}{l}\left(p+1\right)d_n\right\}\sin\frac{n\pi y}{l}\end{array}\right) \tag{A3.10}$$

## LIST OF REFERENCES.

25. The Heaviside function (16.1) includes the sign “$\geq$”. Therefore, if we put $\delta = 0$, the ends and the beginnings of adjoining time segments [see (14.1)]

## FIGURE CAPTIONS

FIG. 1 (a) Schematic drawing of the experimental setup.

(b) The time profile of the sample's heating.

FIG. 2 Typical pyroelectric response of the PNJ.

(a) The periodic excitation response.

$T$=40.2 K; $f$ =733 Hz; $\Delta t$= 0.68 msec; $t_1$=0.06 msec; $t_2$ =0.62 msec.

$P$: (1) 570 mW/cm$^2$, (2) 930 mW/cm$^2$, (3) 330 mW/cm$^2$

(b) The single pulse excitation response.

$T$ = 50.2 K, $\Delta t$ = 1.32 msec, $t_1$ = 0.12 msec, $t_2$ = 1.2 msec;

The duration between pulses is 15 msec.

$P$ : (1) 1020 mW/cm$^2$, (2) 1530 mW/cm$^2$

FIG. 3 (a) The pyroelectric coefficient of the PNJ at different bias voltages.

(b) The "voltage pyroelectric coefficient" (see text) of the PNJ at different bias voltages.

FIG. 4 (a) The temporal variation of the periodic signal at $T$ = 80.1 K, $V_b$=0 and at different chopping frequencies:

(1) $f$ =735 Hz; (2) $f$ =417 Hz; (3) $f$ =109 Hz; (4) $f$ =40 Hz; (5) $f$ =17 Hz.

(b) The corresponding calculated curves

FIG. 5 (a) The temporal variation of the periodic signal at $f$ =40 Hz, $V_b$= 0 and at different temperatures:

(1) T=12.8 K; (2) T =25.2 K; (3) T=60.2 K; (4) T=80.1 K; (5) T=130 K.

(b) The corresponding calculated curves

FIG. 6 The temperature dependence of the periodic signal amplitude at $V_b$=0 and at different frequencies.

$\Delta t$ is the duration of the illumination (or darkness) period.

FIG. 7 The bias voltage dependence of the periodic signal amplitude.

(a) $f$ = 733 Hz and different temperatures;

(b) T=50K and different frequencies.

FIG. 8 The kinetics of the single pulse response (a), and the result of fitting (b) at $T$ = 12.5 K and different durations of illumination ($\Delta t$).

(1) $\Delta t$ = 4.6 msec; (2) $\Delta t$ = 10 msec; (3) $\Delta t$ = 18 msec.

FIG. 9 The steps in the procedure of reconstructing the temperature kinetics in the junction region:

(a) The measured pyro-signal $U(t)$.

(b) $dU/dt$ calculated from data in Fig. 9a.

Data taken at $T$=25.5 K and at different frequencies:

(1) $f$ = 733 Hz, $\Delta t$ =0.68 msec; (2) $f$ = 433 Hz, $\Delta t$ =1.2 msec; (3) $f$ = 109 Hz, $\Delta t$ =4.6 msec; (4) $f$ = 40 Hz, $\Delta t$ =12.6 msec; (5) $f$ = 17Hz, $\Delta t$ =29 msec.

FIG. 10 The steps in the procedure of reconstruction the temperature kinetics in the junction region (continuation from previous FIG.):

(a) $dT/dt$ calculated from the data.

(b) $\Delta T(t)$ obtained by integration of the data in Fig. 10a.

FIG. 11 The temperature dependence of the amplitude of the temperature increment $\Delta T_m$.

FIG. 12 The temporal variation of the currents $I_1$ (black) and $I_2$ (grey-dashed).

$f$ = 40 Hz; $\Delta t$ = 12.5 msec

(a) $T$=12.8 K; (b) $T$=50 K; (c) T=70 K; (d) $T$=90 K; (e) $T$=130 K;

FIG. 13 Single pulse excitation. Reconstruction of $T(t)$ and the kinetics of the displacement, $I_1$ (black), and conduction, $I_2$ (grey-dashed), currents.

T = 60.2 K, P = 1330 mW/cm$^2$, $\Delta$t = 4.6 msec, the dark period is 50 msec.

FIG. 14 (a) The thermal relaxation time $t^*(T, f)$

(b) The heat transfer parameter $G_T(T)$

FIG. 15 The effective heated volume VS (T, f).

(a) For different $\Delta$t (msec): (1) 12.6; (2) 4.6; (3) 1.2

(b) For different temperatures: (1) 11.4 K; (2) 25.2 K; (3) 80 K.

FIG. 16 The parameter $p\ (T, f) = t^*/\tau_e$

(a) For different $\Delta$t (msec):

(1) 29; (2) 12.6; (3) 4.6; (4) 1.2; (5) 0.68

(b) For different temperatures:

(1)11.4 K: (2) 12.8 K; (3) 25.2 K: (4) 80 K; (5) 130 K

Appendix 1.

FIG. A1.1 Schematic drawing of the chopper window area ($D \times H$) and the laser beam area ($L \times H$). The rotation is clockwise. The FIG. shows the instant when the laser beam and the chopper window start to overlap.

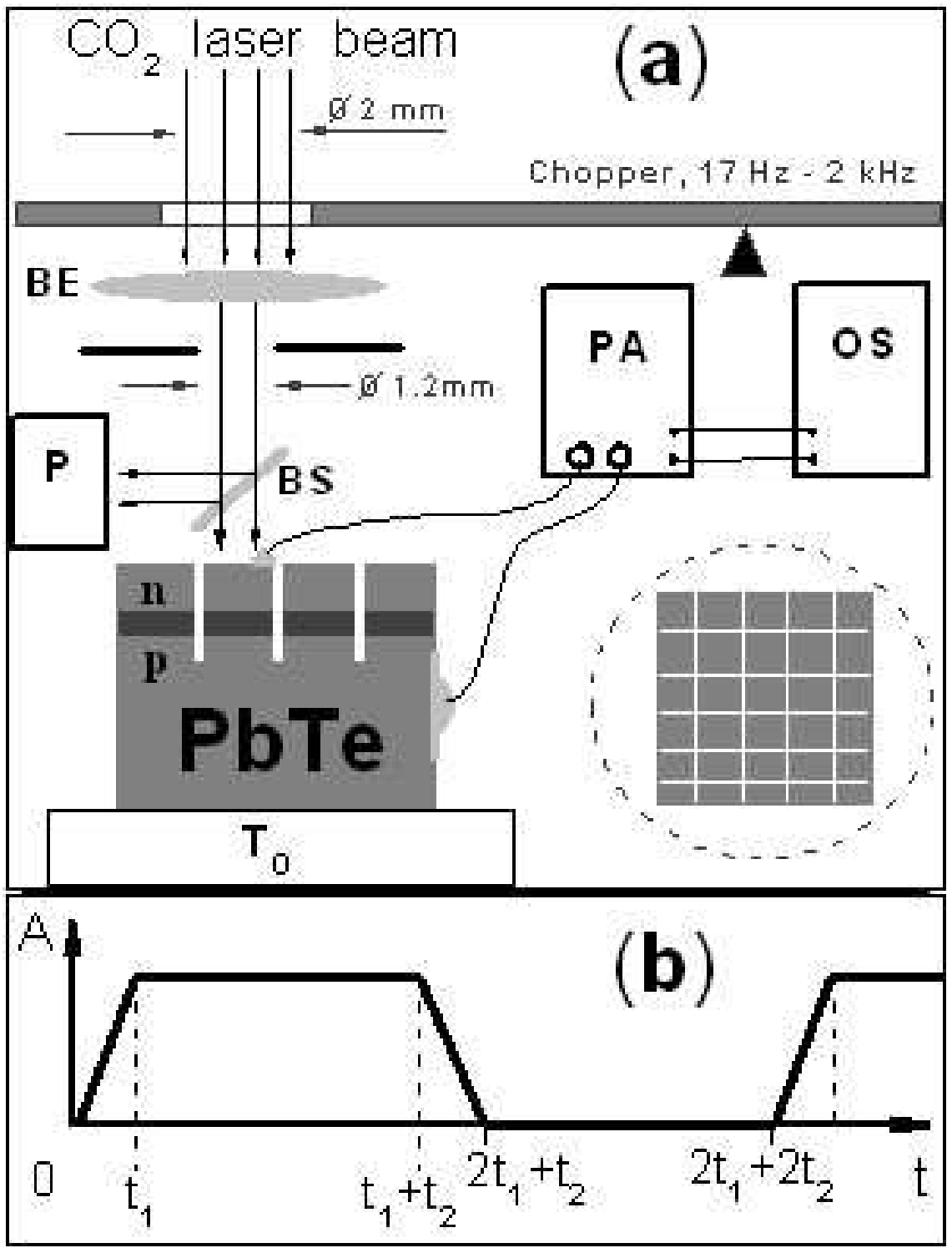

Figure 1

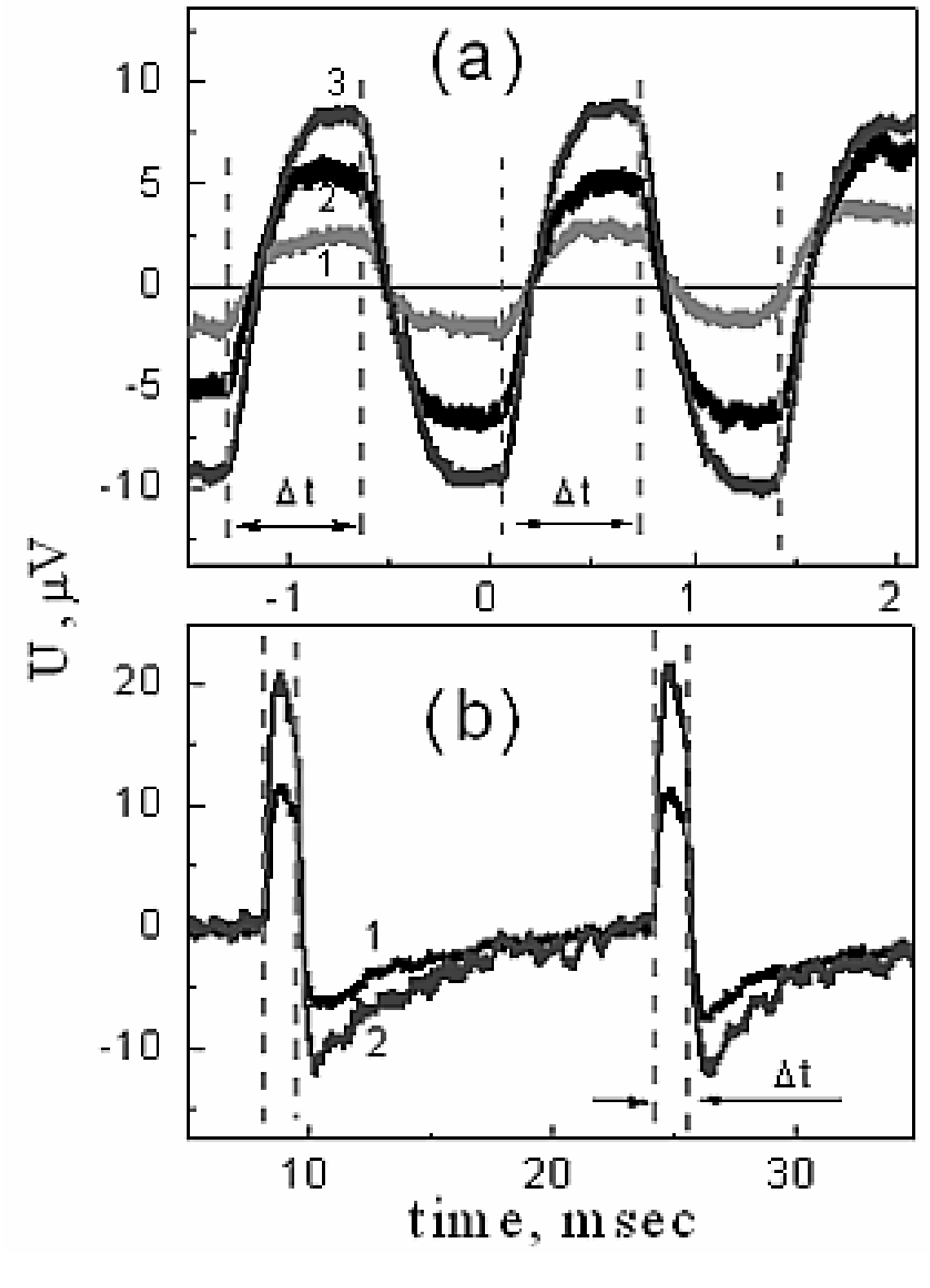

Figure 2

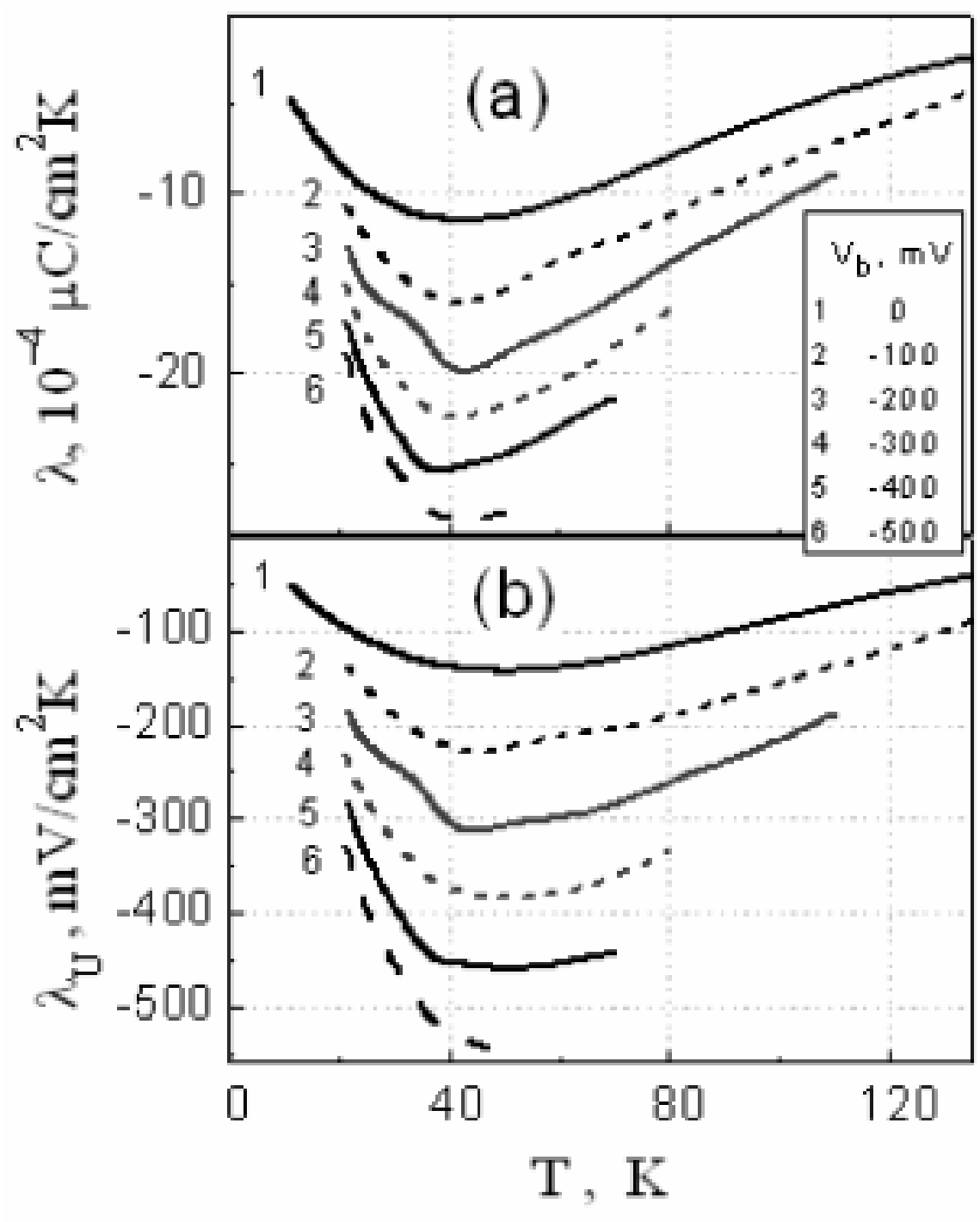

Figure 3

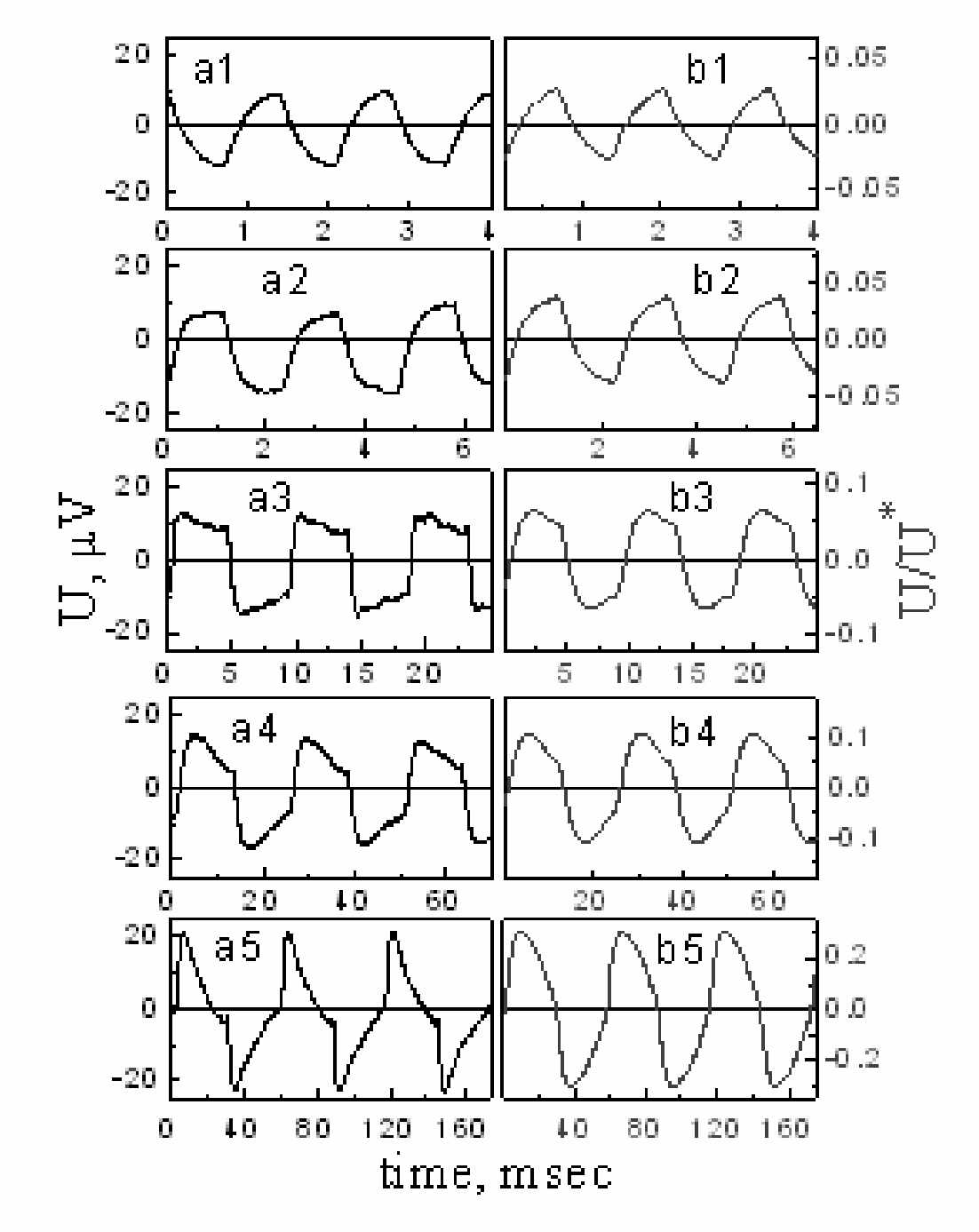

Figure 4

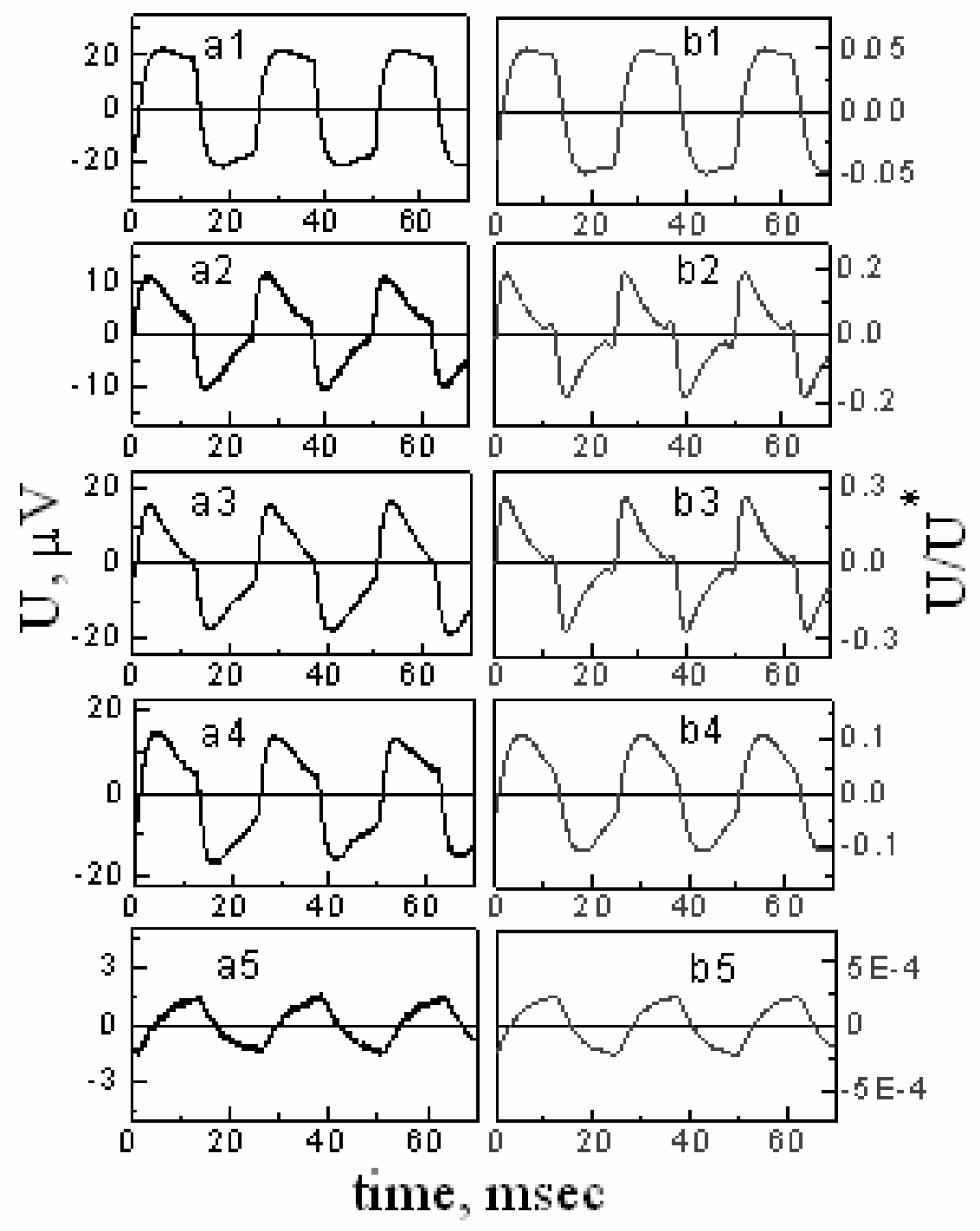

Figure 5

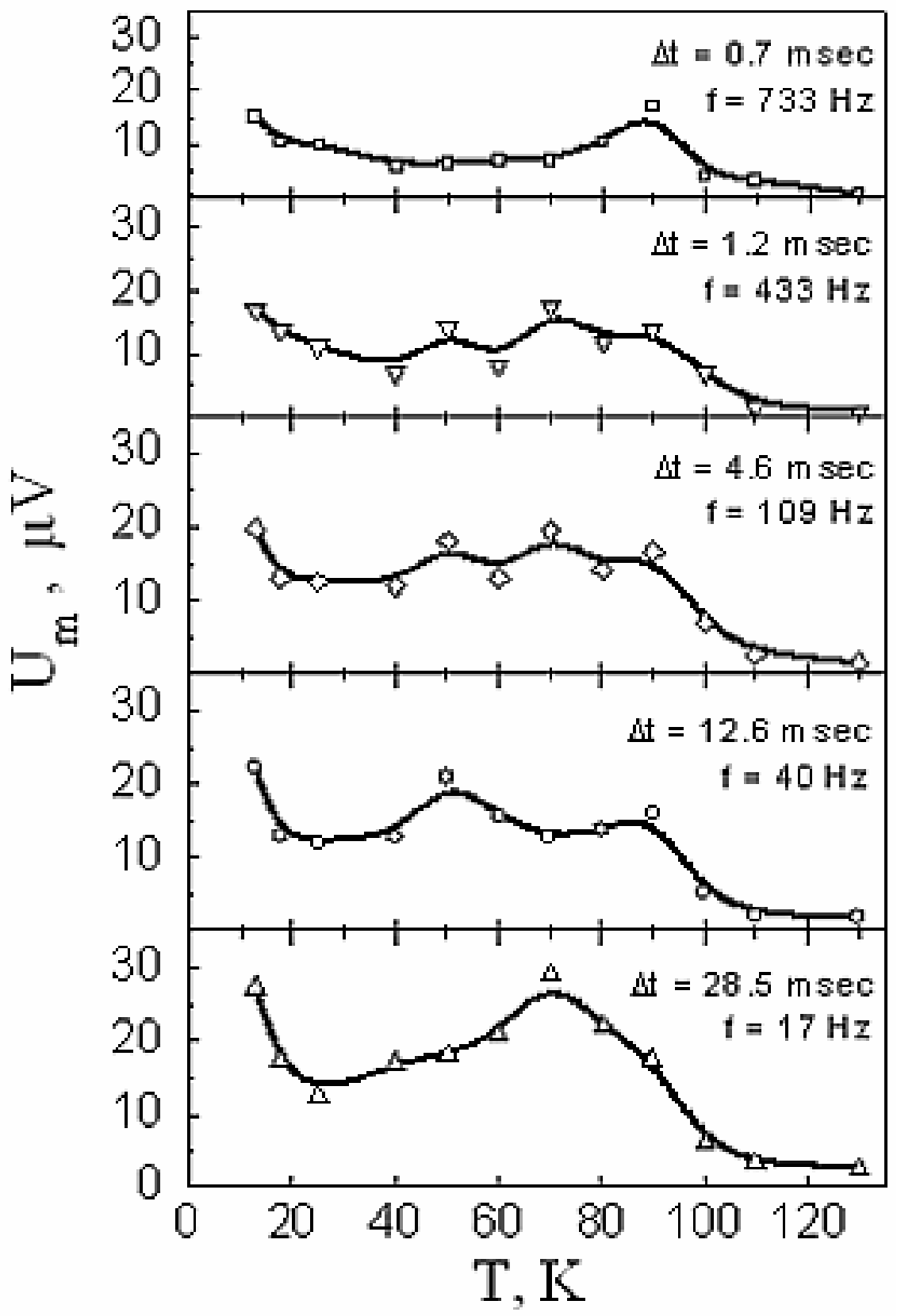

Figure 6

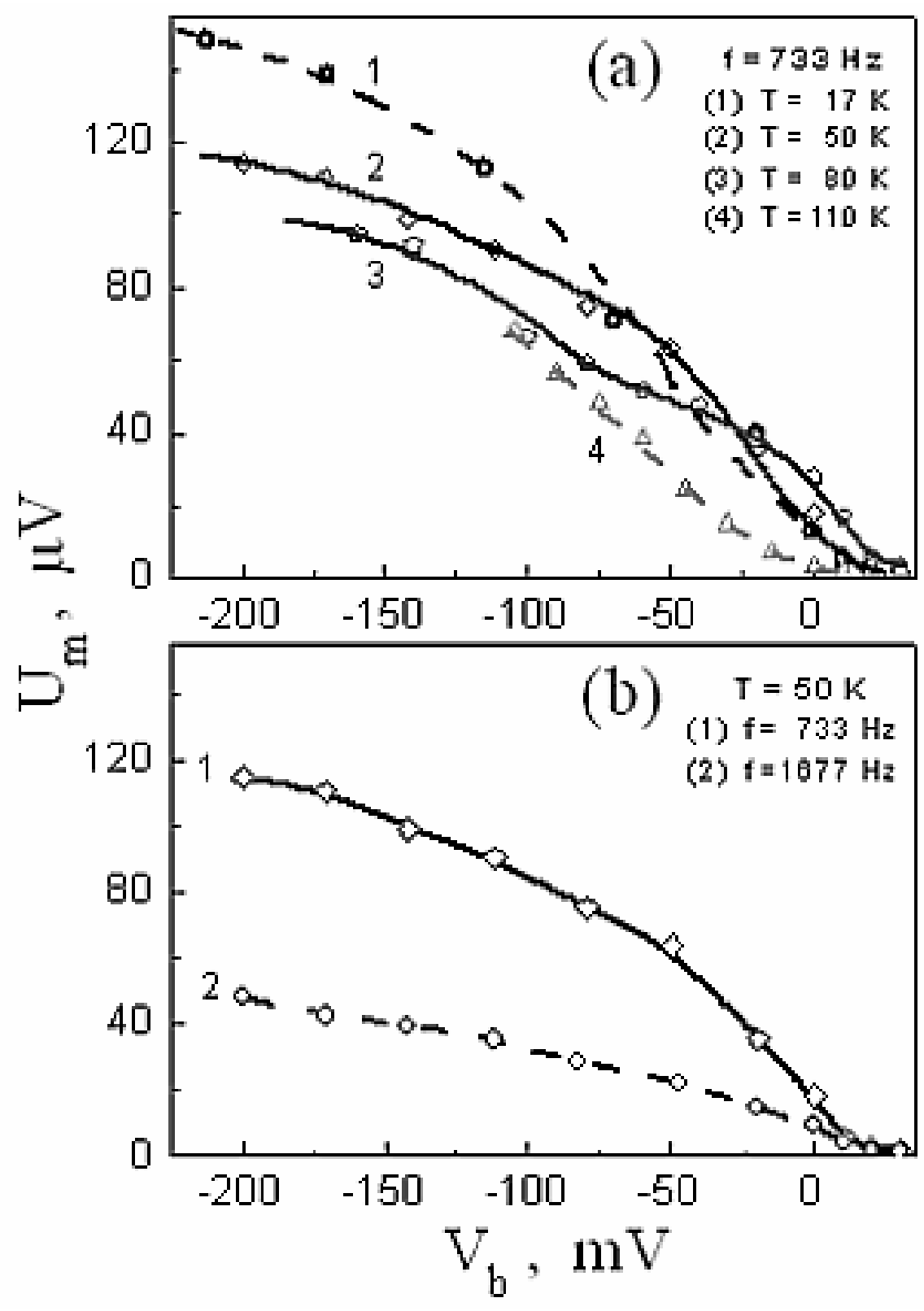

Figure 7

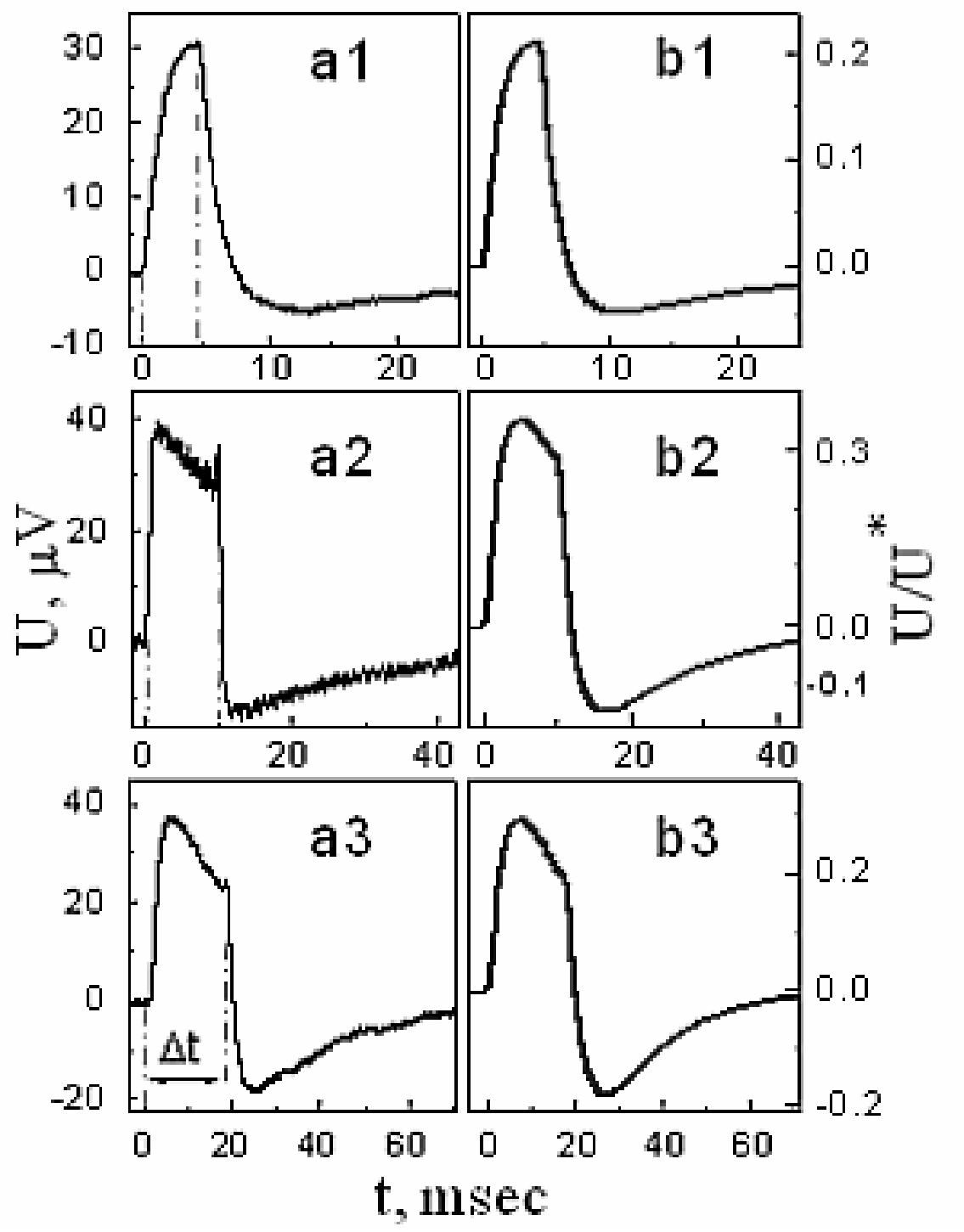

Figure 8

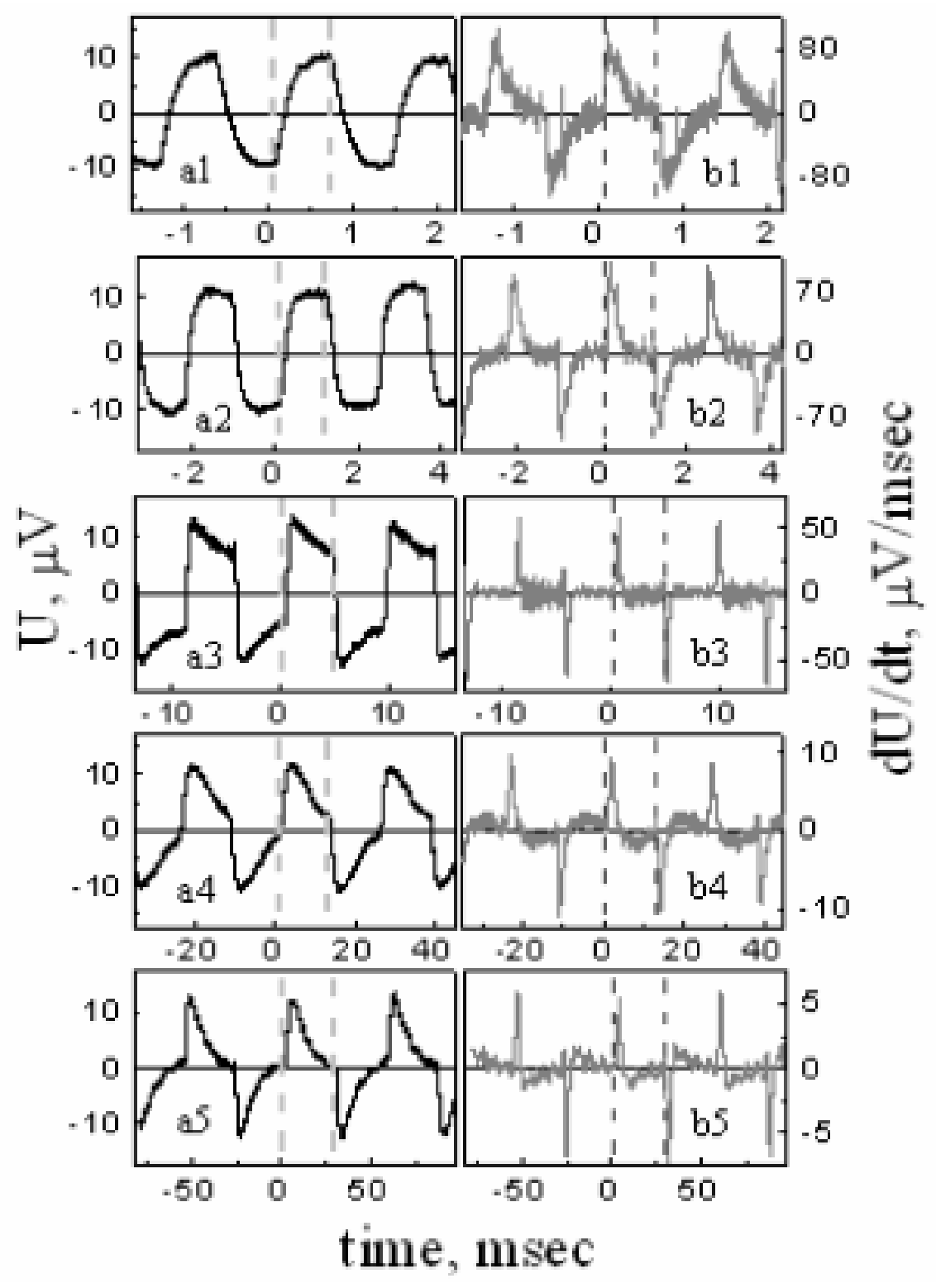

Figure 9

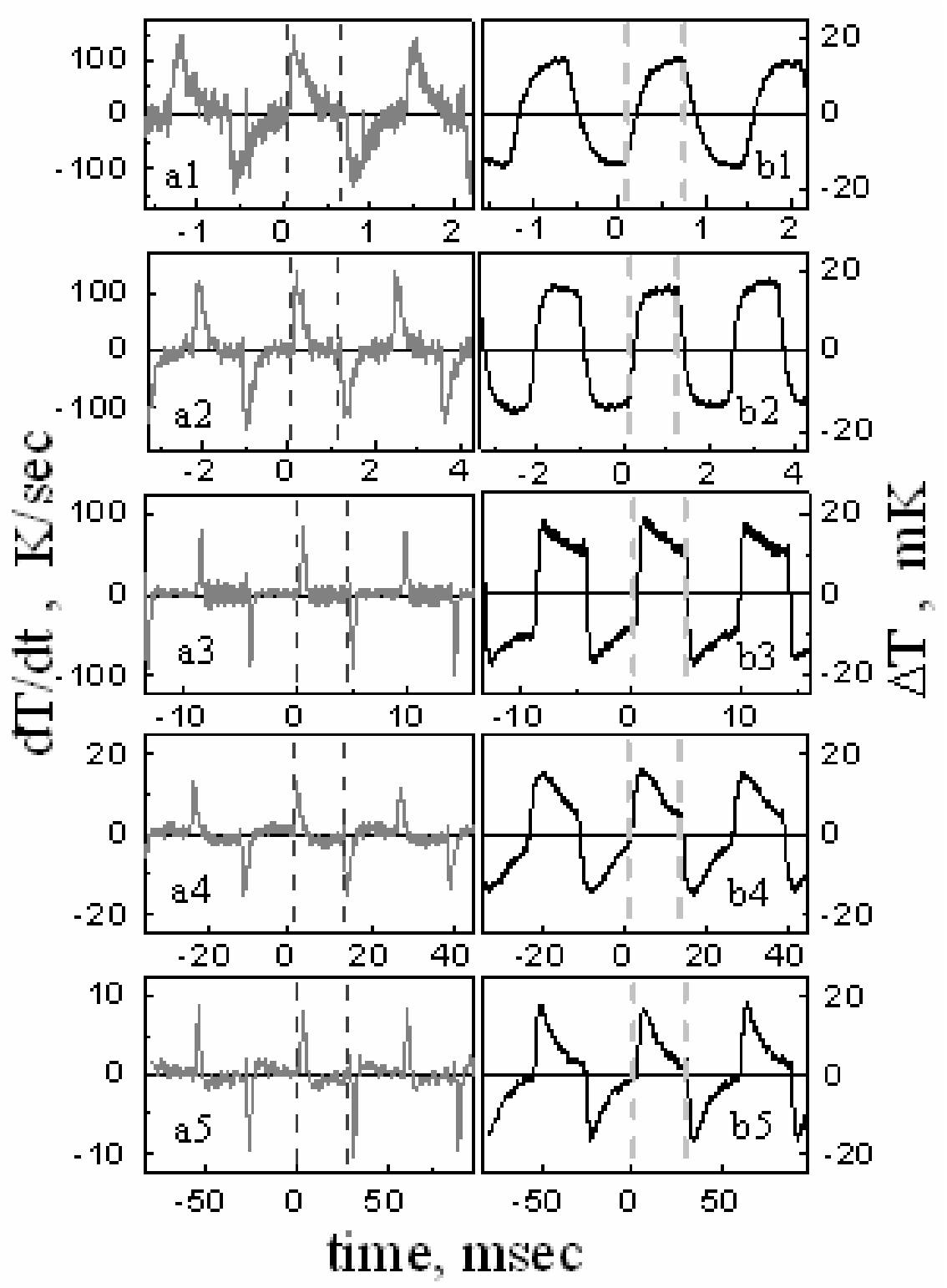

Figure 10

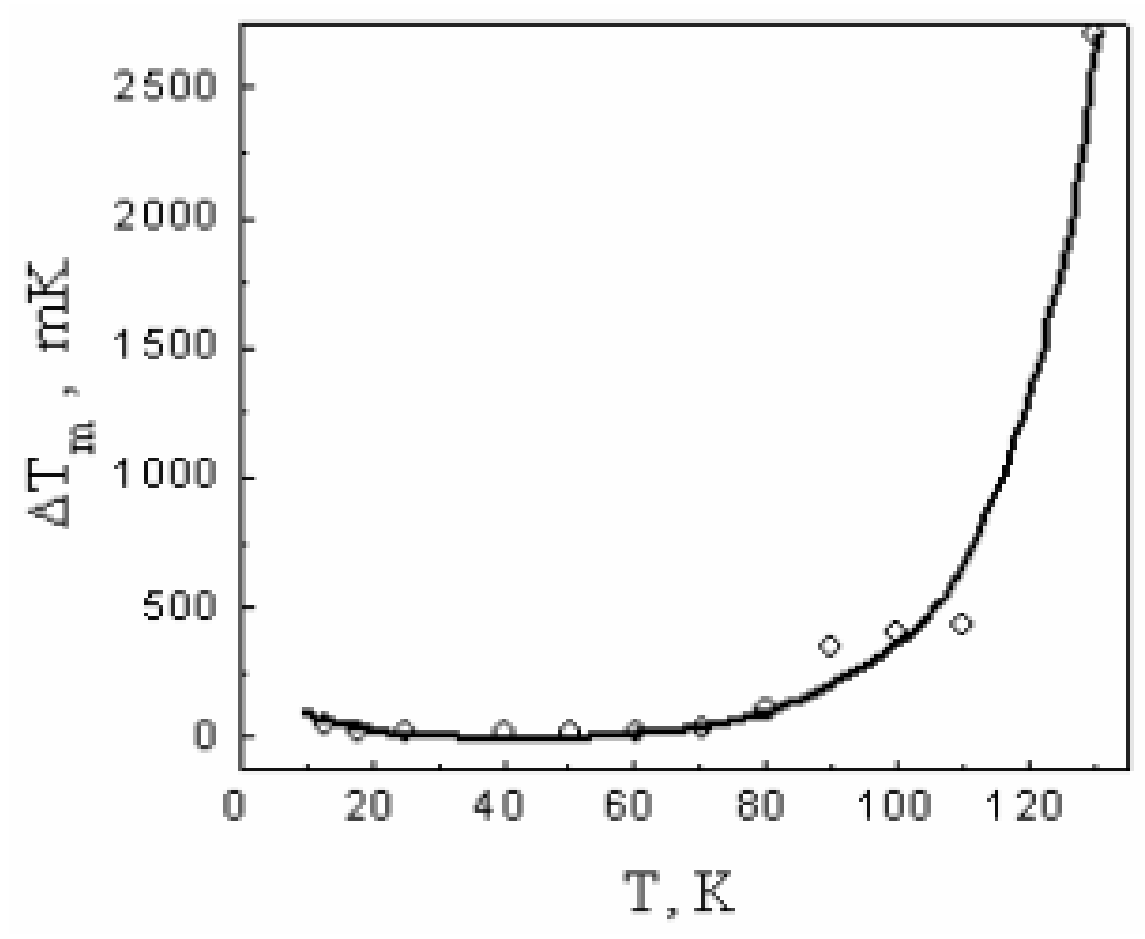

Figure 11

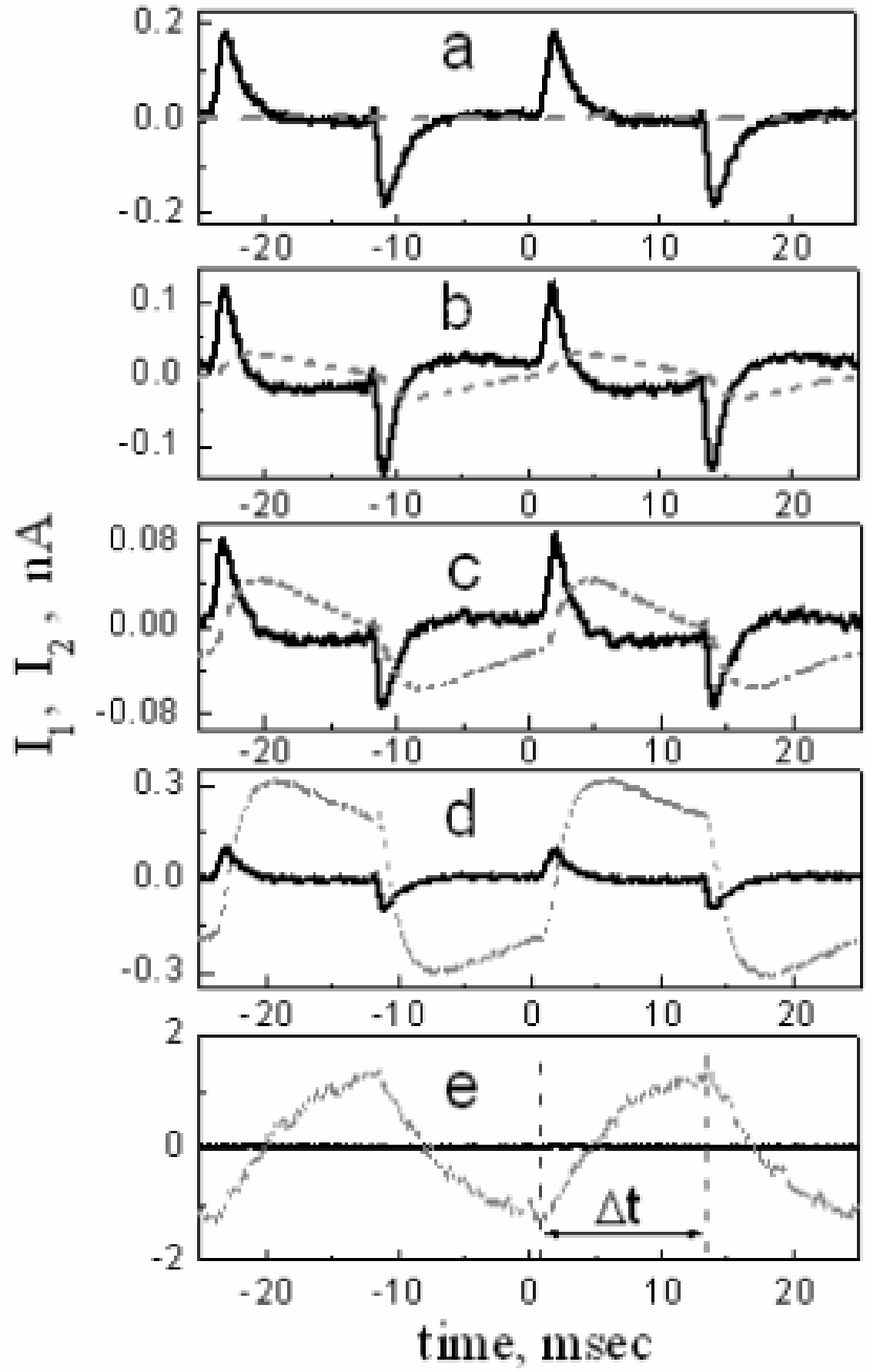

Figure 12

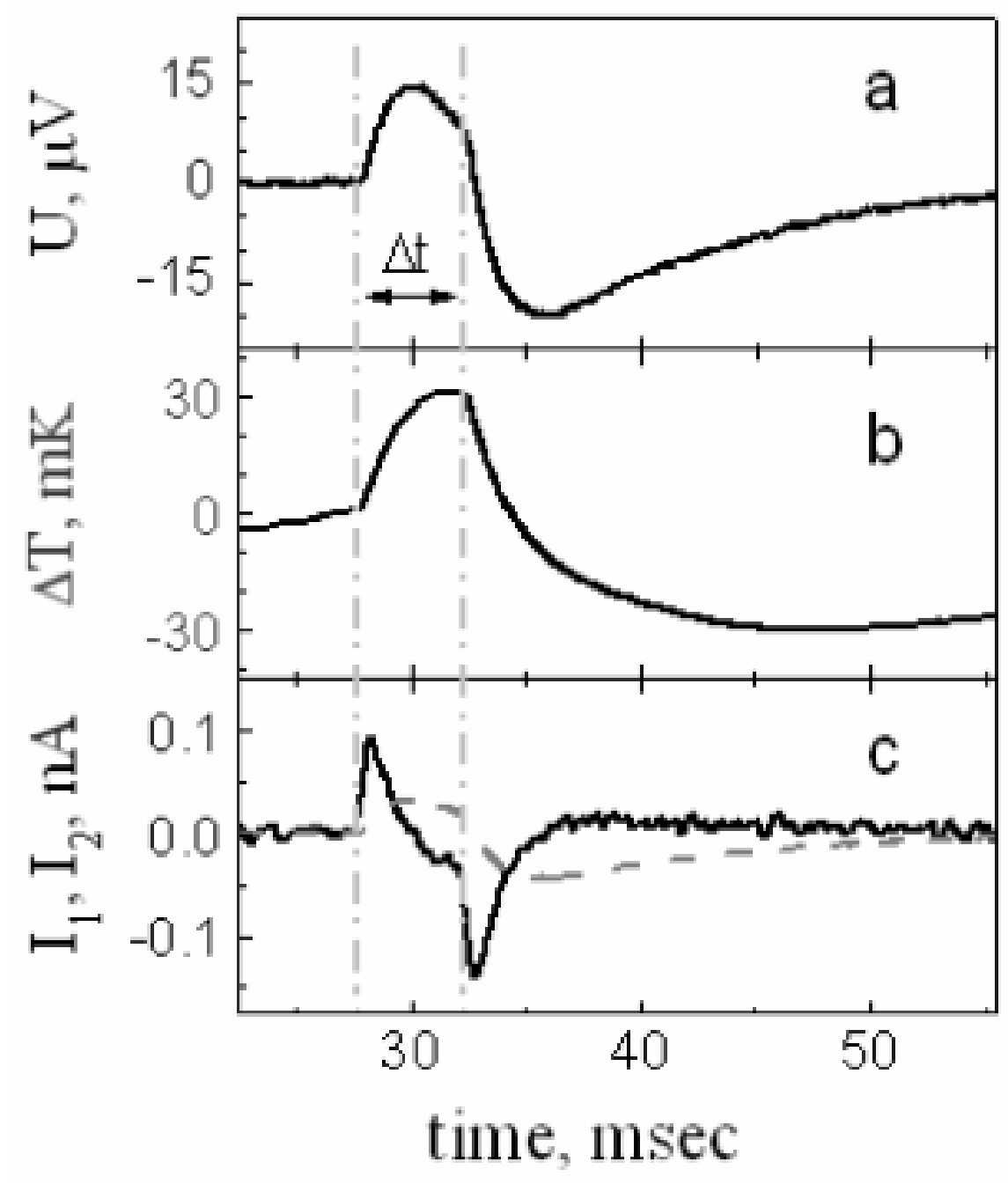

Figure 13

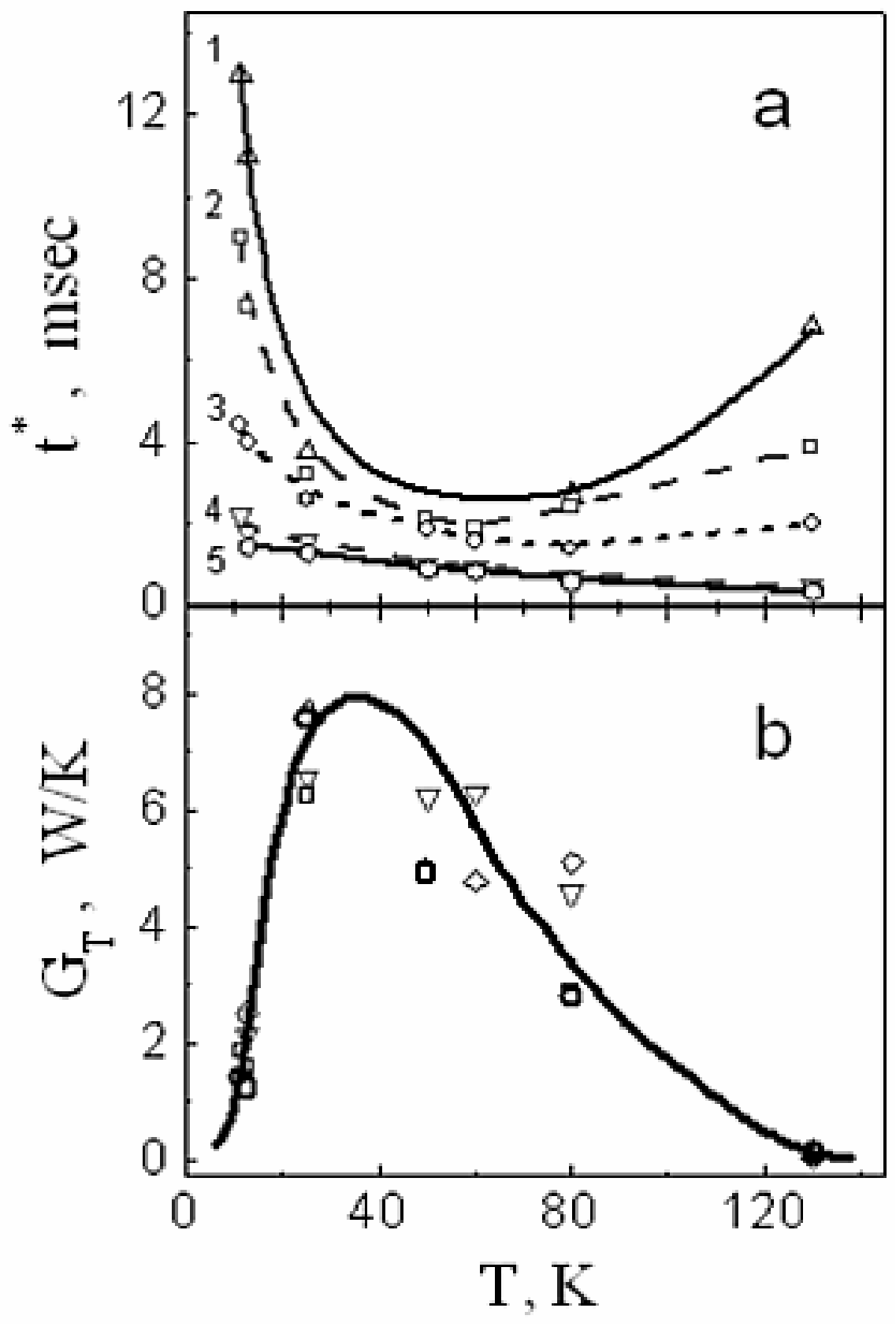

Figure 14

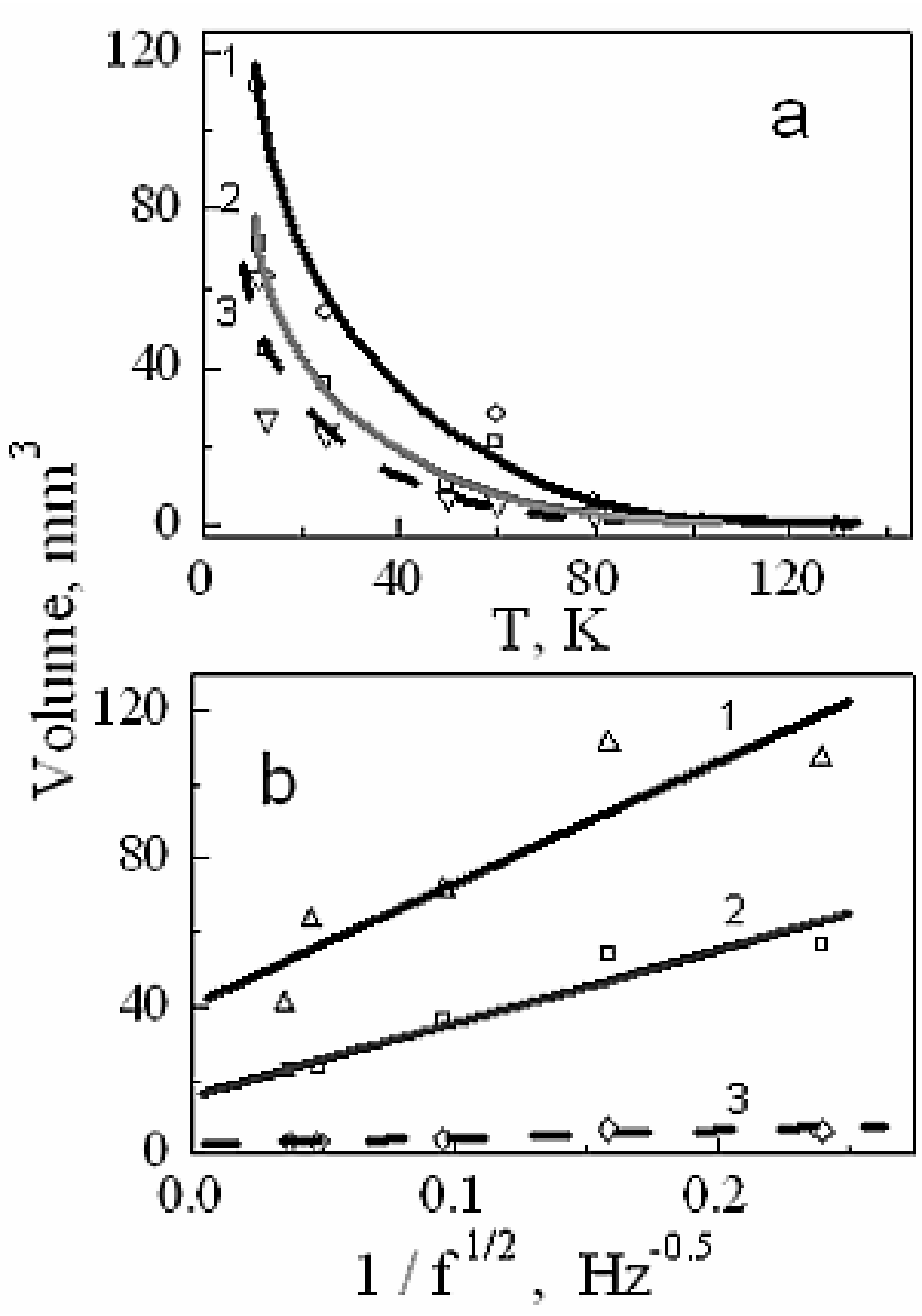

Figure 15

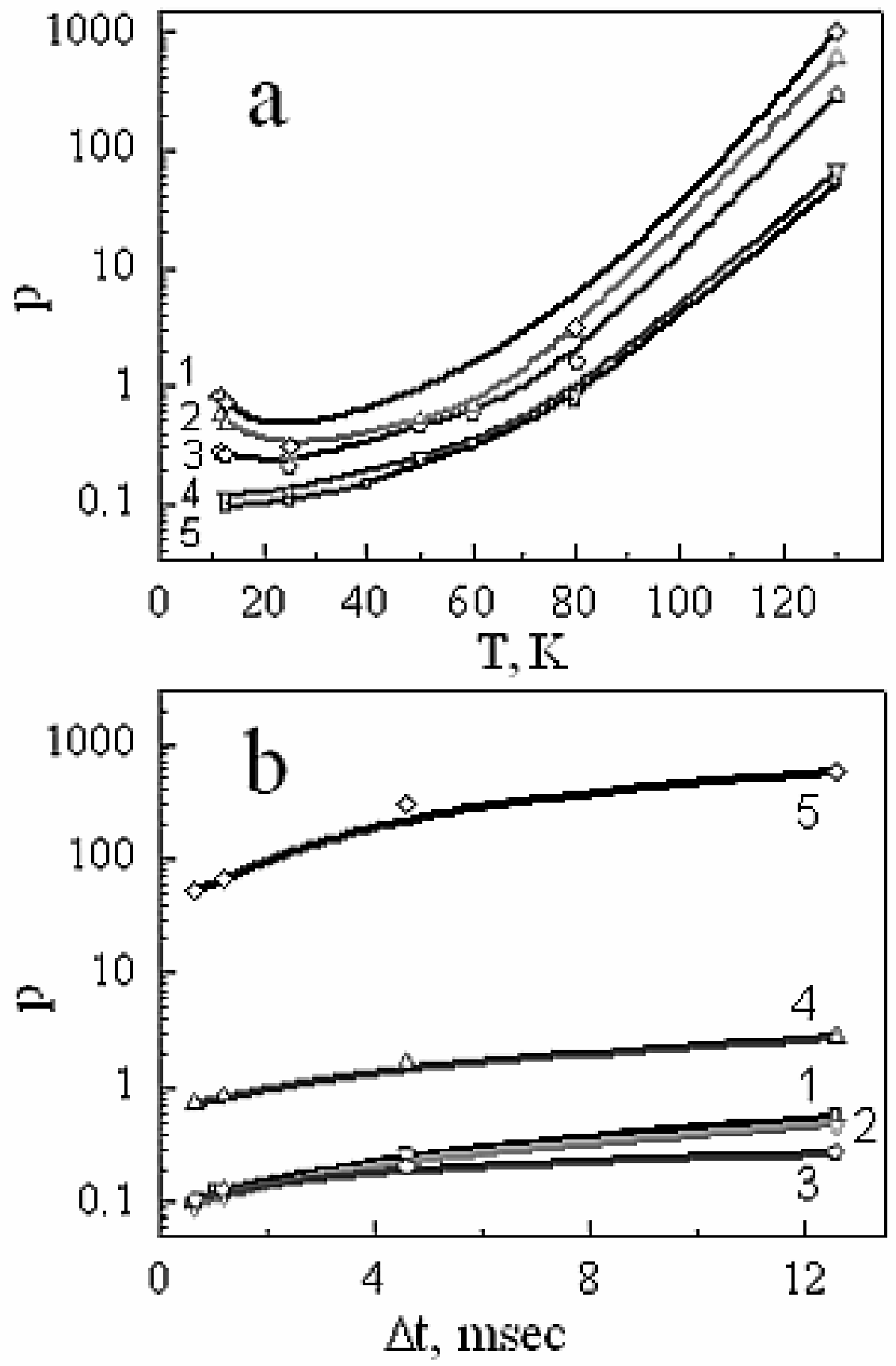

Figure 16

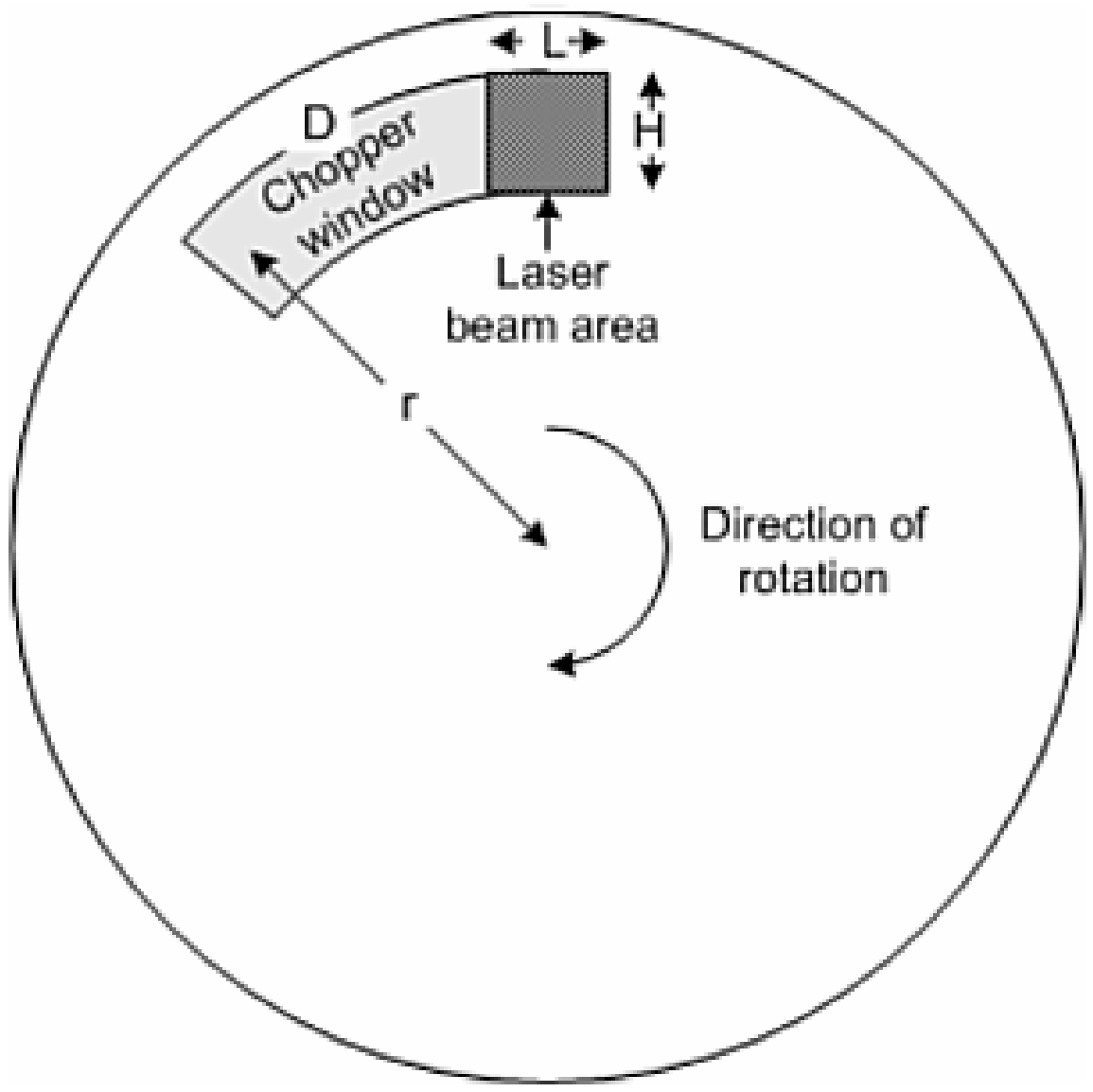

Figure A1